\documentclass[journal,twoside]{IEEEtran}
\usepackage{cite}
\usepackage{amsmath,amssymb,amsfonts}
\usepackage{algorithmic}
\usepackage{algorithm}
\usepackage{graphicx}
\usepackage{textcomp}
\usepackage{url}
\usepackage{amsthm}
\usepackage{tikz}
\usepackage{booktabs}
\usepackage{multirow}
\usepackage{subcaption}
\usepackage{bm}
\usepackage{nomencl}
\usepackage{pgfplots}
\usepgfplotslibrary{groupplots}
\pgfplotsset{compat=1.18}
\makenomenclature

\usetikzlibrary{arrows.meta, positioning, patterns, shapes, automata, calc, decorations.pathreplacing}

\newtheorem{theorem}{Theorem}
\newtheorem{definition}{Definition}
\newtheorem{lemma}{Lemma}
\newtheorem{assumption}{Assumption}
\newtheorem{remark}{Remark}
\newtheorem{proposition}{Proposition}

\newcommand{\R}{\mathbb{R}}
\newcommand{\norm}[1]{\left\lVert#1\right\rVert}

\newcommand{\domain}{\mathcal{D}}
\newcommand{\guard}{\mathcal{G}}
\newcommand{\reset}{\mathcal{R}}

\newcommand{\KL}{\mathcal{KL}}
\newcommand{\K}{\mathcal{K}}
\newcommand{\Etrack}{\mathcal{E}_{\text{track}}}
\newcommand{\Wdist}{\mathcal{W}_{\text{dist}}}

\begin{document}

\title{Hierarchical Preemptive Holistic Collaborative Systems for Embodied Multi-Agent Systems: Framework, Hybrid Stability, and Scalability Analysis}

\author{Ting~Peng, \IEEEmembership{Member,~IEEE}
\thanks{This work was supported in part by the Key Laboratory for Special Area Highway Engineering of Ministry of Education, Chang’an University.}
\thanks{T. Peng is with the Key Laboratory for Special Area Highway Engineering of Ministry of Education, Chang’an University, Xi’an, 710064, China (e-mail: t.peng@ieee.org).}}

\maketitle

\begin{abstract}
The coordination of Embodied Multi-Agent Systems in constrained physical environments requires a rigorous balance between safety, scalability, and efficiency. Traditional decentralized approaches, such as reactive collision avoidance, can succumb to local minima or reciprocal yielding stand-offs due to a lack of future intent awareness. Conversely, centralized planning suffers from intractable computational complexity and single-point-of-failure vulnerabilities. To address these limitations, we propose the \textit{Hierarchical Preemptive Holistic Collaborative (Prollect)} framework. This architecture generalizes the Preemptive Holistic Collaborative System (PHCS) by decomposing the global coordination problem into topologically connected subspace optimizations. We formalize the system as a Hybrid Automaton, introducing a three-stage receding horizon mechanism—comprising frozen execution, preliminary planning, and proactive look-ahead windows—with explicit padding that prevents races between coordination dissemination and intent updates. Crucially, we introduce a robust timing protocol with a mandatory ``Idle Buffer'' that acts as a dwell-time constraint to prevent Zeno behaviors and ensure computational stability under jitter. Furthermore, we formalize a \textit{Shadow Agent} protocol to ensure seamless trajectory consistency across subspace boundaries, treated as an Input-to-State Stability (ISS) problem. We also derive probabilistic safety guarantees under Bernoulli communication dropouts, linking the frozen horizon length to blackout tolerance. Under standard MPC terminal ingredients and a tube/tracking-envelope abstraction, we prove recursive feasibility and value-function decrease. Comparative Monte Carlo simulations against reactive baselines (VO/ORCA) and a DMPC-like replanning baseline show collision-free execution in the tested scenarios, while Prollect improves completion throughput and reduces velocity disruption via proactive look-ahead.
\end{abstract}

\begin{IEEEkeywords}
Multi-agent systems, distributed control, hybrid systems, receding horizon control, preemptive holistic collaboration, boundary handover, Lyapunov stability.
\end{IEEEkeywords}

\section*{Nomenclature}
\addcontentsline{toc}{section}{Nomenclature}
\begin{IEEEdescription}[\IEEEusemathlabelsep\IEEEsetlabelwidth{$W_1, W_2, W_3$}]
\item[$\mathcal{EA}$] Set of $N$ agents $\{a_1, \dots, a_N\}$.
\item[$\mathcal{W}$] Global workspace $\mathcal{W} \subset \R^n$.
\item[$\mathcal{W}_i$] The $i$-th topological subspace.
\item[$\mathcal{N}_i$] Logical coordinator for subspace $\mathcal{W}_i$.
\item[$\mathcal{B}_k(x)$] Occupied volume of agent $k$ at state $x$.
\item[$\Pi_{\text{safe}}(\cdot)$] Shared discrete-time safety projection operator applied to intended commands.
\item[$v^{int}(t_k)$] Intended velocity command at update time $t_k$ (before projection).
\item[$v^{exec}(t_k)$] Executed velocity command at update time $t_k$ (after projection).
\item[$\eta$] Numerical threshold used to decide whether the projection modified the intended command.
\item[$t_{step}$] Control update cycle duration.
\item[$t_{adj}$] Computation time for trajectory optimization and conflict resolution within $t_{step}$.
\item[$t_{tx}$] Communication time to transmit trajectories/consensus to agents.
\item[$t_{pad}$] Padding/guard time separating coordination from intent updates (require $t_{pad}>t_{tx}$).
\item[$t_{frozen}$] Frozen-window duration (immutable commitment horizon).
\item[$t_{planning}$] Planning-window duration (active optimization horizon).
\item[$t_{lookahead}$] Look-ahead duration used for proactive conflict detection/negotiation.
\item[$W_1, W_2, W_3$] Frozen, Planning, and Look-ahead windows.
\item[$\mathcal{ST}_k$] Spatiotemporal tube of agent $k$.
\item[$\mathcal{S}_i$] Set of Shadow Agents (boundary crossing).
\item[$J_i(\cdot)$] Holistic cost function for subspace $i$.
\item[$V(\cdot)$] Lyapunov candidate function.
\item[$\mathcal{X}_f$] Terminal invariant set.
\item[$Q\succeq 0,R\succ 0$] Stage-cost weight matrices (tracking and control effort).
\item[$\delta_{safe}>0$] Required minimum separation margin in the tightened (tube-inflated) safety constraint.
\item[$\alpha$] Reliability multiplier for frozen window sizing.
\item[ProjAct] Projection activation rate: fraction of control calls where the shared safety projection modifies the intended command.
\item[$\mathcal{H}$] Hybrid Automaton tuple.
\end{IEEEdescription}

\section{Introduction}

\IEEEPARstart{T}{he} deployment of large-scale multi-agent systems (MAS) in critical infrastructure—such as autonomous highway systems, automated warehousing, and search-and-rescue swarms—demands control strategies that are not only efficient but provably safe. A specific and challenging class of these systems involves \textit{Multi-Embodied Agents} (MEA). Unlike theoretical point-mass agents often studied in consensus literature, MEAs possess non-trivial physical volumes, complex geometries, and kinematic constraints. As noted by Pfeifer and Bongard \cite{Pfeifer2006Body}, the physical embodiment fundamentally shapes the interaction dynamics, introducing non-convex geometric constraints that render standard potential-field methods insufficient.

The complexity of coordinating MEAs arises from the dual requirement of satisfying local dynamic constraints while adhering to global safety constraints (collision avoidance). As the density of agents increases, the free configuration space becomes increasingly disconnected, leading to the well-known ``freezing robot problem'' or livelock scenarios where purely reactive agents become trapped in local minima, unable to negotiate right-of-way without high-level coordination.

\subsection{Motivation and Gap Analysis}
Current approaches to MAS coordination face significant trade-offs:
\begin{enumerate}
    \item \textbf{Scalability vs. Optimality:} Centralized planners (e.g., MAPF) scale exponentially ($O(c^N)$), making them unsuitable for fleets larger than a few dozen agents. Decentralized reactive methods (e.g., ORCA) scale linearly ($O(N)$) but sacrifice optimality and deadlock freedom.
    \item \textbf{Theory vs. Practice:} Many theoretical Distributed MPC (DMPC) papers assume perfect, synchronous communication. In real-world robotic networks, packet loss, bandwidth limits, and computational jitter are pervasive. A control theory that ignores these often fails in deployment.
    \item \textbf{Boundary Consistency:} In hierarchical or partitioned systems, the "handover" problem—safely transferring control of an agent from one logical coordinator to another without discontinuity—is often glossed over.
\end{enumerate}

To address these challenges, we build upon the foundational work of Li \textit{et al.} \cite{Li2024Preemptive}, who introduced the \textbf{Preemptive Holistic Collaborative System (PHCS)}. PHCS emphasized "Preemption" (resolving conflicts before they become immediate threats) and "Holism" (collective optimization). However, the original formulation lacked a rigorous control-theoretic stability proof and a mechanism for hierarchical scaling.

\subsection{Contributions}
In this paper, we extend PHCS into the \textbf{Hierarchical Prollect Framework}. The term "Prollect" denotes \textit{Preemptive}, \textit{Holistic}, and \textit{Collaborative}. Our specific contributions are:

1) \textbf{Hybrid Automaton Formalism:} We model the coordinator dynamics as a Hybrid Automaton. We define a robust timing protocol with an explicitly defined ``Idle Buffer'' ($t_{step} > 1.5 t_{adj}$). We prove that this buffer acts as a minimum dwell time, ensuring the system avoids Zeno instability and is robust to computational jitter.

2) \textbf{Hierarchical Decomposition with Shadow Agents:} We decompose the global workspace into topological subspaces. We introduce a ``Shadow Agent'' protocol that allows adjacent coordinators to maintain consensus on agents crossing boundaries without a central authority. We provide a complexity analysis showing this reduces the problem size from global $O(N^3)$ to local $O(N_{local}^3)$.

3) \textbf{Rigorous Stability Proofs:} Leveraging MPC theory \cite{Rawlings2017MPC} and consensus principles \cite{OlfatiSaber2007Consensus}, we provide detailed proofs for Recursive Feasibility (Theorem \ref{thm:rf}) and Asymptotic Stability (Theorem \ref{thm:asym}). Unlike previous sketches, we explicitly construct the candidate trajectories using terminal invariant sets and bound the cost decrease.

4) \textbf{Comprehensive Validation:} We benchmark the Prollect framework against reactive baselines (VO-projection and ORCA \cite{VanDenBerg2011ORCA}) and a distributed MPC-style baseline (DMPC-BR: iterative best-response MPC), and include ablations that isolate the effect of preemptive look-ahead. Results show collision-free execution and reduced velocity disruption due to proactive conflict resolution.

\noindent\textbf{What Prollect adds beyond the shared safety projection.}
To make attribution explicit, all evaluated methods apply the same discrete-time safety projection layer that enforces short-horizon separation at the command level. Prollect's contribution is therefore \emph{not} ``having a safety filter,'' but rather:
(i) higher throughput/completion relative to purely reactive baselines under symmetry stand-offs,
(ii) smaller average velocity disruption $\overline{\Delta v}$ relative to nominal goal-seeking,
(iii) bounded preemption rate (preemptive adjustments are triggered rarely), and
(iv) reduced reliance on the safety projection, quantified by a \emph{projection activation rate} (fraction of control calls where the projection changes the intended command).

\medskip
\noindent In particular, in the intersection benchmark (Table~\ref{tab:intersection_eff}), Prollect achieves 100\% completion with a low preemption rate while reducing velocity disruption and substantially reducing ProjAct relative to the reactive VO-projection baseline, indicating that the coordinator resolves conflicts \emph{before} the reactive safety layer is forced to intervene. Compared to distributed replanning (DMPC-BR), Prollect attains similar completion with smaller velocity disruption via sparse, proactive adjustments.

\section{Related Work}

The problem of multi-agent coordination has been studied extensively. We categorize existing literature into three streams to contextualize the Prollect framework.

\subsection{Centralized Planning}
Centralized approaches treat the multi-agent system as a single, high-dimensional robot. Techniques like Coupled $A^*$ or Multi-Agent Path Finding (MAPF) operate on a discretized grid. While Conflict-Based Search (CBS) and its variants have improved scalability, they remain NP-hard in the worst case. Furthermore, centralized planners are single points of failure; if the central server disconnects, the entire fleet halts. In contrast, the Prollect framework is inherently distributed; the failure of one subspace coordinator only affects agents within that local region.

\subsection{Decentralized Reactive Control}
Decentralized methods compute control inputs based solely on local sensing. Velocity Obstacles (VO) and Optimal Reciprocal Collision Avoidance (ORCA) create local constraints in velocity space that aim to avoid imminent collisions over a finite look-ahead horizon under their modeling assumptions \cite{Fiorini1998VO,VanDenBerg2011ORCA}. Artificial Potential Fields (APF) use attractive and repulsive forces. While computationally efficient ($O(N)$), these methods are ``myopic'': they do not explicitly coordinate future intent and can exhibit reciprocal yielding, oscillations, or non-termination in dense, topologically constrained environments. Prollect addresses this by incorporating a Look-ahead Window ($W_3$) specifically designed to detect and resolve such conflicts preemptively.

\medskip
\noindent In particular, VO \cite{Fiorini1998VO} and ORCA \cite{VanDenBerg2011ORCA} provide strong short-horizon collision-avoidance behavior but do not generally guarantee deadlock freedom in dense, topologically constrained environments, motivating the proactive, windowed mechanism in Prollect.

\medskip
\noindent\textbf{Safety filters and barrier certificates.}
An alternative to purely geometric reactive avoidance is to enforce safety via online constraint enforcement, such as control barrier functions (CBFs) and related safety-filter formulations \cite{Ames2019CBF,Wabersich2021PCBF}. These methods typically compute a minimally-modified safe input (often via a quadratic program) that keeps the state within a forward-invariant safe set. In this paper, we adopt a shared discrete-time safety projection layer to match sampled execution and to isolate Prollect's contribution beyond last-moment safety correction.
Formal safety verification/control for intersection collision avoidance is also studied in the TAC literature; see, e.g., \cite{Ahn2018IntersectionSafety}.

\subsection{Distributed Model Predictive Control (DMPC)}
DMPC is the closest relative to our work. Agents solve local optimization problems and exchange trajectories with neighbors to reach a Nash equilibrium.
\begin{itemize}
    \item \textbf{Serial DMPC:} Agents optimize sequentially. This avoids conflicts but introduces significant latency, scaling linearly with the number of neighbors.
    \item \textbf{Parallel DMPC:} Agents optimize simultaneously. This requires iterative consensus rounds to converge, which can be bandwidth-intensive.
    \item \textbf{Tube-based DMPC:} Addresses uncertainty by optimizing a nominal trajectory surrounded by a ``tube'' of invariant sets.
\end{itemize}
Standard DMPC assumes that the computation time is negligible compared to the control step. In reality, solving non-convex trajectory optimization problems is computationally expensive. Our Prollect framework explicitly accounts for this via the ``Idle Buffer'' constraint, treating computation time as a state in a hybrid system, ensuring practical stability.

\medskip
\noindent Recent work continues to refine scalable non-centralized MPC and partitioning strategies; see, e.g., \cite{Riccardi2025Partitioning}. For distributed MPC under communication imperfections and inexact (dual/consensus) optimization, see representative Automatica treatments such as \cite{Kohler2019InexactDualDMPC,Li2021DMPCCommNoise}. Recent TAC papers further develop distributed MPC formulations and separable/ADMM-based decompositions for multiagent networks; see, e.g., \cite{Shorinwa2024SeparableDMPC,Mallick2025SwitchADMM}.

\medskip
\noindent\textbf{MPC-based avoidance in multi-robot systems.}
Beyond purely reactive avoidance, MPC-style formulations are widely used to incorporate mission objectives while enforcing safety/avoidance constraints; see, e.g., avoidance-feature MPC and adaptive collision-avoidance navigation in \cite{AlvesDoSanto2024AvoidMPC,Verginis2021AdaptiveAvoid}.
Related stochastic MPC treatments for collision avoidance in cooperating autonomous platforms also appear in the recent IEEE literature; see, e.g., \cite{Cao2025SMPCAvoid}.
For stochastic multi-agent safety/containment under uncertainty using tube-based MPC ideas, see, e.g., \cite{Li2022TubeContainment}. Conceptually, this is aligned with our tube/tracking-envelope bridge (Assumption~\ref{assump:track}): rather than claiming perfect tracking of the nominal plan, we reason about safety by enforcing tightened constraints on an inflated set (here represented by $\mathcal{B}_k(\tau)\oplus\Etrack$) that upper-bounds executed motion.

\section{Problem Formulation}

\subsection{Embodied Agent Dynamics}
Consider a set of $N$ agents $\mathcal{EA} = \{a_1, \dots, a_N\}$ operating in a global workspace $\mathcal{W} \subset \R^n$ ($n \in \{2,3\}$). We assume the agents are subject to non-holonomic kinematic constraints (e.g., unicycle model).

\begin{definition}[Unicycle Embodied Agent]
Each agent $a_k$ is defined by the state vector $x_k = [p_x, p_y, \theta]^T \in \mathcal{X}_k \subset \R^3$ and control input $u_k = [v, \omega]^T \in \mathcal{U}_k \subset \R^2$. The dynamics are:
\begin{equation}
    \dot{x}_k(t) = \begin{bmatrix} \cos(\theta_k) & 0 \\ \sin(\theta_k) & 0 \\ 0 & 1 \end{bmatrix} u_k(t)
    \label{eq:unicycle}
\end{equation}
The physical volume is denoted by $\mathcal{B}_k(x_k)$, a compact subset of $\R^2$ representing the footprint of the robot at state $x_k$.
\end{definition}

\begin{assumption}[Boundedness]
The state constraint set $\mathcal{X}_k$ and control constraint set $\mathcal{U}_k$ are compact and contain the origin. The workspace $\mathcal{W}$ is compact and convex.
\end{assumption}

\noindent We rewrite \eqref{eq:unicycle} in control-affine form:
\begin{equation}
\begin{aligned}
\dot{x}_k &= g_1(x_k)\, v_k + g_2(x_k)\, \omega_k,\\
g_1(x) &= \begin{bmatrix}\cos\theta\\ \sin\theta\\ 0\end{bmatrix},\qquad
g_2(x) = \begin{bmatrix}0\\ 0\\ 1\end{bmatrix}.
\end{aligned}
\label{eq:control-affine}
\end{equation}

\begin{assumption}[Regularity]
For each $k$, the vector fields $g_1, g_2$ are locally Lipschitz on $\mathcal{X}_k$, and there exist constants $L_f, M_f > 0$ such that for all $x_1,x_2\in\mathcal{X}_k$ and $u\in\mathcal{U}_k$,
\begin{equation}
\begin{aligned}
\norm{f_k(x_1,u)-f_k(x_2,u)} &\le L_f\norm{x_1-x_2},\\
\norm{f_k(x,u)} &\le M_f.
\end{aligned}
\end{equation}
\end{assumption}

\begin{proposition}[Lie Bracket and STLC]
The unicycle system \eqref{eq:control-affine} is small-time locally controllable at any $x\in\mathcal{X}_k$.
\end{proposition}
\begin{proof}
The Lie bracket of $g_1$ and $g_2$ is $[g_1,g_2]=\frac{\partial g_2}{\partial x}g_1-\frac{\partial g_1}{\partial x}g_2=\begin{bmatrix}\sin\theta\\ -\cos\theta\\ 0\end{bmatrix}$. The set $\{g_1(x),g_2(x),[g_1,g_2](x)\}$ spans $\R^3$ for all $\theta$. By Chow--Rashevskii (bracket-generating) theorem, the system is STLC.
\end{proof}

\subsection{Planning model versus execution model (used in proofs and simulations)}
\label{subsec:model_mismatch}
The paper presents unicycle dynamics to capture embodiment and nonholonomy. However, the coordination layer in Prollect operates on a \emph{planning abstraction} that is intentionally simpler: each agent submits an intended planar velocity command and executes the projected command in discrete time, while low-level tracking closes the gap to the true robot dynamics.
Formally, the safety and feasibility analysis is stated for the tube-inflated footprint model (Assumption \ref{assump:track} and Proposition \ref{prop:tube_safety}), where the planner enforces separation on $\mathcal{B}_k(\tau)\oplus\Etrack$. In simulation, we implement this abstraction as planar velocity integration with a shared discrete-time safety projection layer $\Pi_{\text{safe}}$ and log ProjAct to quantify how often this last-moment correction is required. This is the intended ``theory--implementation bridge'' used throughout Sections VI--VIII.

\subsection{Hierarchical Subspace Decomposition}
To ensure scalability, we partition $\mathcal{W}$ into $M$ topological subspaces $\{\mathcal{W}_i\}_{i=1}^M$ such that $\mathcal{W} = \bigcup_{i} \mathcal{W}_i$.

\begin{assumption}[Subspace Connectivity]
The subspace graph $G_{sub} = (\mathcal{V}_{sub}, \mathcal{E}_{sub})$ is connected. Adjacent subspaces $\mathcal{W}_i$ and $\mathcal{W}_j$ share a boundary region $\partial \mathcal{W}_{ij}$ with non-zero measure, allowing for safe handover.
\end{assumption}

A logical coordinator $\mathcal{N}_i$ is assigned to each $\mathcal{W}_i$. At any time $t$, agent $a_k$ is ``owned'' by $\mathcal{N}_i$ if its centroid lies within $\mathcal{W}_i$. If $\mathcal{B}_k(x_k) \cap \mathcal{W}_j \neq \emptyset$, $a_k$ is instantiated as a \textit{Shadow Agent} in $\mathcal{N}_j$.

\subsection{Communication Topology and Graph Laplacian}
The hierarchical architecture induces a coordinator-level communication graph $G_{comm}=(\mathcal{V}_{comm},\mathcal{E}_{comm})$:
\begin{equation}
\begin{aligned}
\mathcal{V}_{comm} &= \{\mathcal{N}_1,\dots,\mathcal{N}_M\},\\
(\mathcal{N}_i,\mathcal{N}_j)\in\mathcal{E}_{comm} &\iff \partial\mathcal{W}_{ij}\neq\emptyset.
\end{aligned}
\end{equation}
Let $\gamma_{ij}=\gamma_{ji}>0$ denote a coupling weight if $(i,j)\in\mathcal{E}_{comm}$ and $\gamma_{ij}=0$ otherwise. The (weighted) Laplacian $L\in\R^{M\times M}$ is
\begin{equation}
L_{ij}=\begin{cases}
-\gamma_{ij}, & i\neq j\\
\sum_{\ell\neq i}\gamma_{i\ell}, & i=j.
\end{cases}
\end{equation}

\begin{assumption}[Connectivity]
The graph $G_{comm}$ is connected; equivalently, the algebraic connectivity satisfies $\lambda_2(L)>0$.
\end{assumption}

\noindent This Laplacian enters the boundary-consensus coupling induced by Shadow Agents: neighboring coordinators penalize disagreement on the shared (shadowed) state/trajectory, yielding a standard consensus-like contraction term governed by $\lambda_2(L)$ \cite{OlfatiSaber2007Consensus}.

\section{The Prollect Coordination Framework}

The Prollect framework transforms the continuous-time coordination problem into a discrete-event system modeled as a Hybrid Automaton.

\subsection{Robust Three-Stage Receding Horizon}
We partition the look-ahead horizon $T$ into three functional intervals.

\begin{figure*}[t]
\centering
\resizebox{0.95\textwidth}{!}{%
\begin{tikzpicture}[>=Stealth, node distance=0.5cm]
    \draw[->, thick] (-0.5,0) -- (14.8,0) node[right] {Time $\tau$};
    \coordinate (t0) at (0,0);
    \coordinate (tf) at (5.0,0);  
    \coordinate (tp) at (7.0,0);  
    \coordinate (T) at (14.0,0);  
    
    \fill[gray!20] (t0) rectangle (5.0, 0.8);
    \draw[thick] (t0) rectangle (5.0, 0.8);
    \node at (2.5, 0.4) {\textbf{$W_1$: Frozen}};
    
    \fill[blue!10] (5.0, 0) rectangle (7.0, 0.8);
    \draw[thick] (5.0, 0) rectangle (7.0, 0.8);
    \node at (6.0, 0.4) {\textbf{$W_2$: Plan}};
    
    \fill[green!10] (7.0, 0) rectangle (14.0, 0.8);
    \draw[thick] (7.0, 0) rectangle (14.0, 0.8);
    \node at (10.5, 0.4) {\textbf{$W_3$: Look-ahead}};
    
    \draw[|<->|, thick, blue] (0, 2.5) -- (1.5, 2.5) node[midway, fill=white] {$t_{step}$};
    \node[blue, font=\footnotesize, above] at (0.75, 2.7) {Cycle};
    
    \draw[fill=red!30] (0, 1.3) rectangle (0.8, 1.8);
    \node[font=\scriptsize, align=center] at (0.4, 1.55) {$t_{adj}$};
    
    \draw[<->, dashed] (0.8, 1.55) -- (1.5, 1.55);
    \node[font=\scriptsize, above] at (1.15, 1.55) {Idle};
    
    \draw[fill=yellow!30] (1.5, 1.3) rectangle (2.0, 1.8);
    \node[font=\scriptsize, align=center] at (1.75, 1.55) {Tx (bg)};
    
    \draw[<->, thick] (0, -0.5) -- (5.0, -0.5) node[midway, fill=white] {$t_{frozen}$};
    
    \node[align=left, draw=black, thin, rounded corners, fill=white, font=\scriptsize] at (10.2, 2.0) {
        \textbf{Timing Constraints:}\\
        1. $t_{step} > 1.5 \times t_{adj}$ (Dwell time)\\
        2. $t_{frozen}=\alpha t_{step}$, $\alpha\ge 1$ (Robustness)\\
        3. $t_{pad}>t_{tx}$ (No intent/coordination race)
    };
\end{tikzpicture}%
}
\caption{The Prollect temporal structure. The mandatory \textit{Idle} buffer enforces dwell time for computation. Transmission (Tx) is asynchronous/background and can be pipelined across cycles; $t_{pad}>t_{tx}$ provides a logical separation between coordination dissemination and intent updates in $W_3$.}
\label{fig:horizon}
\end{figure*}
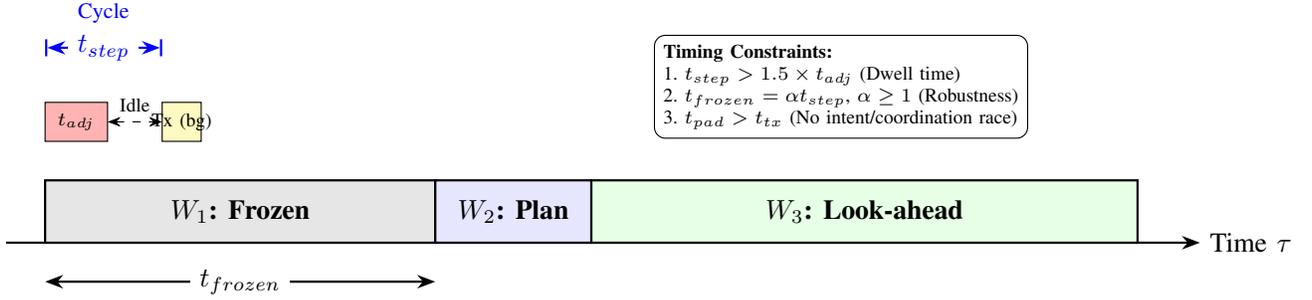

\begin{definition}[Functional Windows]
\begin{enumerate}
    \item \textbf{Frozen Window ($W_1$):} The interval $[t_k,\, t_k + t_{frozen}]$. Trajectories here are immutable commitments. This ensures that even if communication fails for multiple cycles (up to $t_{frozen}/t_{step}$), agents have a valid, collision-free path to execute.
    \item \textbf{Planning Window ($W_2$):} The interval $(t_k + t_{frozen},\, t_k + t_{frozen} + t_{planning}]$. This is the active optimization domain. Holistic adjustments are calculated here to optimize flow and energy.
    \item \textbf{Look-ahead Window ($W_3$):} The interval $(t_k + t_{frozen} + t_{planning},\, t_k + T]$. We split it into (i) a short \emph{coordination padding} interval $(t_k + t_{frozen} + t_{planning},\; t_k + t_{frozen} + t_{planning} + t_{pad}]$ and (ii) an \emph{intent-modification} interval $(t_k + t_{frozen} + t_{planning} + t_{pad},\; t_k + T]$.
    The padding is chosen to avoid any race between coordinator coordination/consensus dissemination and agents modifying intents: require
    \begin{equation}
        t_{pad} > t_{tx},
    \end{equation}
    where $t_{tx}$ (sometimes denoted $t_{Tx}$) is the transmission time. Transmission is executed in the background (asynchronous), so the padding is a \emph{logical} separation ensuring that intent changes only begin after the coordinator has finished publishing the relevant coordination information for the cycle.
\end{enumerate}
\end{definition}

\begin{remark}[Recommended ``Graceful'' Tuning]
Based on system design analysis, we recommend the following ratio for graceful operation under uncertainty:
\begin{equation}
t_{frozen}:t_{plan}:t_{look}:t_{step}:t_{tx}:t_{pad} \approx 10:5:10:1:0.5:3.
\end{equation}
This configuration provides: (1) high robustness to communication outages ($t_{frozen} \gg t_{step}$), (2) sufficient maneuver duration ($t_{plan}$), (3) deep foresight for conflict resolution ($t_{look}$), and (4) race-free synchronization ($t_{pad} \gg t_{tx}$). While this implies a long optimization horizon ($W_2 \cup W_3$), the computational load is kept feasible ($t_{adj} < t_{step}$) by the hierarchical decomposition, which bounds the local agent count $N_i$ regardless of global scale.
\end{remark}

\begin{remark}[Intent submission and satisfaction]
If agents submit their intents to the coordinators sufficiently early (so they are visible within the intent-modification subinterval of $W_3$), Prollect can proactively reconcile them by making minor trajectory/velocity adjustments while preserving safety. When the set of submitted intents is jointly feasible under dynamics and hard safety constraints, the coordinator can satisfy all agents' needs simultaneously; otherwise, Prollect returns a best-effort compromise via the holistic cost/priority weights.

\medskip
\noindent\textbf{No-conflict (snapshot) rule.} To eliminate conflicts between ``modifying'' and ``coordinating'' actions, the coordinator consumes a \emph{snapshot} of intents at a well-defined cut-off time, and any later intent updates are buffered for the next cycle (double-buffering/versioning). The padding $t_{pad}>t_{tx}$ ensures that agents only start modifying intents after the coordinator has disseminated the coordination information they should react to.
\end{remark}

\begin{algorithm}
\caption{Coordinator Cycle (Prollect Preemptive Timing)}
\begin{algorithmic}[1]
\STATE \textbf{Given:} cycle start $t_k$, windows $W_1,W_2,W_3$, padding $t_{pad}>t_{tx}$
\STATE \textbf{Intent snapshot:} read buffered intents at cut-off $\tau=t_k+t_{frozen}+t_{planning}+t_{pad}$
\STATE \textbf{Detect:} predict conflicts in $(t_k+t_{frozen}+t_{planning},\, t_k+t_{frozen}+t_{planning}+t_{step}]$
\IF{conflict detected}
    \STATE \textbf{Preempt:} adjust actions over $(t_k+t_{frozen},\, t_k+t_{frozen}+t_{planning}+t_{step}]$ (minor $\Delta v$)
\ENDIF
\STATE \textbf{Optimize:} solve local HVP over $W_2\cup W_3$ with tightened safety (inflated by $\Etrack$)
\STATE \textbf{Publish:} asynchronously transmit consensus/trajectories (background; delivery budget $t_{tx}$)
\STATE \textbf{Execute:} agents follow the immutable prefix on $W_1$; intent updates occur only after the padding
\end{algorithmic}
\end{algorithm}

\subsection{Coordinator as a Hybrid Automaton}
To rigorously analyze the timing constraints and prove the absence of Zeno behavior, we model the coordinator $\mathcal{N}_i$ as a Hybrid Automaton $\mathcal{H}_i = (Q, X, f, \domain, \guard, \reset)$.

\medskip
\noindent This modeling choice follows standard hybrid systems practice \cite{Goebel2012Hybrid}, where dwell-time constraints are a classical tool to preclude Zeno executions and to obtain robustness to timing perturbations.

\begin{itemize}
    \item \textbf{Discrete States $Q$:} $\{q_{calc}, q_{idle}\}$.
    \item \textbf{Continuous State $X$:} A timer $\tau \in \R_{\geq 0}$.
    \item \textbf{Flow $f$:} $\dot{\tau} = 1$ in all states.
    \item \textbf{Domains $\domain$:}
    \begin{itemize}
        \item $\domain(q_{calc}) = \{ \tau : \tau \le t_{adj}^{max} \}$
        \item $\domain(q_{idle}) = \{ \tau : \tau \le t_{step} \}$
    \end{itemize}
    \item \textbf{Guards $\guard$:}
    \begin{itemize}
        \item $q_{calc} \to q_{idle}$: Guard condition is ``Optimization Converged''.
        \item $q_{idle} \to q_{calc}$: Guard condition is $\tau \geq t_{step}$.
    \end{itemize}
    \item \textbf{Reset $\reset$:} $\tau := 0$ upon transition $q_{idle} \to q_{calc}$.
\end{itemize}

\medskip
\noindent\textbf{Asynchronous transmission (background).}
The communication/transmission task does not need to be a blocking discrete phase. In practice, transmission can be executed by a background thread/processor and can overlap with both $q_{calc}$ and $q_{idle}$ (e.g., pipelining: sending the previous cycle's consensus while computing the next). We therefore do \emph{not} require $t_{tx}\ll t_{step}$, nor do we include $t_{tx}$ in the dwell-time calculation below.
Instead, $t_{tx}$ is treated as a delivery latency budget for disseminating the computed consensus/trajectory once available.
A practical engineering requirement is $t_{frozen} \gg t_{tx}$ (including jitter/outage margins), so agents can safely execute the already-committed Frozen Window while messages are in transit and do not depend on instantaneous delivery.

\begin{theorem}[Zeno-Freeness and Dwell Time]\label{thm:zeno}
Let $t_{adj}^{max}$ be the worst-case execution time (WCET) of the optimization solver. If
\begin{equation}
t_{step} > 1.5 \, t_{adj}^{max},
\end{equation}
then the system enforces a minimum dwell time in $q_{idle}$ given by
\begin{equation}
\tau_{dwell} \ge t_{step} - t_{adj}^{max} \ge 0.5\, t_{adj}^{max} > 0,
\end{equation}
and is Zeno-free.
\end{theorem}
\begin{proof}
See Appendix \ref{app:zeno}.
\end{proof}

\section{Distributed Holistic Optimization}

\subsection{Communication Protocol: Spatiotemporal Tubes}
To minimize bandwidth while maintaining safety, agents do not transmit raw state trajectories. Instead, they transmit \textit{Spatiotemporal Tubes}.

\begin{definition}[Spatiotemporal Tube]
For an agent $a_k$, the tube $\mathcal{ST}_k$ is defined as the Minkowski sum of the embodied volume and a tracking error set $\Etrack$:
\begin{equation}
    \mathcal{ST}_k = \bigcup_{\tau \in W_2 \cup W_3} \left( \mathcal{B}_k(x_k(\tau)) \oplus \Etrack \right) \times \{\tau\}
\end{equation}
\end{definition}

This formulation decouples the high-level planning from the low-level tracking control. As long as the low-level controller maintains the agent within $\Etrack$, safety is guaranteed by the planner.

\begin{proposition}[Tube inflation implies physical safety]\label{prop:tube_safety}
Suppose the executed footprint satisfies $\mathcal{B}_k^{exec}(\tau)\subseteq \mathcal{B}_k(\tau)\oplus \Etrack$ for all $\tau$ (tracking bound), and the planner enforces the tightened separation constraint
\[
d\big(\mathcal{B}_k(\tau)\oplus \Etrack,\; \mathcal{B}_j(\tau)\oplus \Etrack\big)\ge \delta_{safe},\ \forall k\neq j.
\]
Then $d(\mathcal{B}_k^{exec}(\tau),\mathcal{B}_j^{exec}(\tau))\ge \delta_{safe}$ for all $\tau$.
\end{proposition}
\begin{proof}
By set inclusion, $\mathcal{B}_k^{exec}(\tau)\subseteq \mathcal{B}_k(\tau)\oplus \Etrack$ and $\mathcal{B}_j^{exec}(\tau)\subseteq \mathcal{B}_j(\tau)\oplus \Etrack$. Distances between subsets are lower-bounded by distances between supersets, yielding the claim.
\end{proof}

\begin{assumption}[Tracking envelope]\label{assump:track}
For each agent $a_k$, the low-level tracking controller and actuation/sampling implement a bounded error envelope: there exists a known compact set $\Etrack$ such that the executed footprint satisfies $\mathcal{B}_k^{exec}(\tau)\subseteq \mathcal{B}_k(\tau)\oplus \Etrack$ for all $\tau\in W_1\cup W_2\cup W_3$ whenever the planned trajectory $\mathcal{B}_k(\tau)$ is followed.
\end{assumption}

\subsection{Shadow Agent Protocol}
A critical innovation of Prollect is the \textbf{Shadow Agent} mechanism for handling boundary conditions.

\begin{algorithm}
\caption{Shadow Agent Handover Protocol}
\begin{algorithmic}[1]
\STATE \textbf{Coordinator $\mathcal{N}_i$ (Sender):}
\FOR{each $a_k \in \mathcal{EA}_i$}
    \STATE Calculate tube $\mathcal{ST}_k$ for $W_2 \cup W_3$
    \FOR{each neighbor $\mathcal{N}_j \in \text{adj}(\mathcal{N}_i)$}
        \IF{$\mathcal{ST}_k \cap \mathcal{W}_j \neq \emptyset$}
            \STATE Serialize $\mathcal{ST}_k$ and transmit to $\mathcal{N}_j$
        \ENDIF
    \ENDFOR
\ENDFOR
\STATE \textbf{Coordinator $\mathcal{N}_j$ (Receiver):}
\STATE Receive $\mathcal{ST}_k$ from $\mathcal{N}_i$
\STATE Instantiate Shadow Agent $a_{k}^{shadow}$
\STATE Add constraint: $\mathcal{B}_m(\tau) \cap \mathcal{ST}_k(\tau) = \emptyset, \forall a_m \in \mathcal{EA}_j$
\end{algorithmic}
\end{algorithm}

\subsection{The Holistic Variational Problem (HVP)}
The optimization problem solved by $\mathcal{N}_i$ is defined as:
\begin{align}
\min_{\mathbf{u}_i(\cdot)} J_i &= \sum_{a_k \in \mathcal{EA}_i} \int_{W_2 \cup W_3} \Big( \|x_k - \sigma_k^{ref}\|_Q^2 + \|u_k\|_R^2 \Big) d\tau \nonumber \\
&\quad + \sum_{a_k \in \mathcal{S}_i} \int_{W_2 \cup W_3} \lambda_{b} \|x_k - \hat{x}_k^{(j)}\|^2 d\tau \nonumber\\
&\quad + \sum_{a_k \in \mathcal{EA}_i} V_f\!\left(x_k(t_k+T)\right)
\label{eq:cost}
\end{align}

\medskip
\noindent\textbf{Stage cost notation.}
For subsequent MPC stability arguments, define the per-agent stage cost
\begin{equation}
\ell_k(x_k,u_k) := \|x_k-\sigma_k^{ref}\|_Q^2 + \|u_k\|_R^2,
\end{equation}
and the coordinator-level stage cost
\begin{equation}
\ell_i(\mathbf{x}_i,\mathbf{u}_i) := \sum_{a_k\in\mathcal{EA}_i}\ell_k(x_k,u_k) + \sum_{a_k\in\mathcal{S}_i}\lambda_b\|x_k-\hat{x}_k^{(j)}\|^2.
\end{equation}

\noindent Here $\sigma_k^{ref}(\tau)$ encodes the agent's \emph{intent} (e.g., goal-reaching reference, preferred progress schedule, or other mission objectives) as submitted to the coordinator. Because intents can be updated in $W_3$, the reference $\sigma_k^{ref}$ can be revised proactively before execution-critical portions enter $W_2$ and $W_1$.

\noindent \textbf{Subject to:}
\begin{enumerate}
    \item \textbf{Dynamics:} $\dot{x}_k = f_k(x_k, u_k), \forall \tau \in W_2 \cup W_3$.
    \item \textbf{Continuity:} $x_k(t_k + t_{frozen}^+) = x_k^{prev}(t_k + t_{frozen}^-)$, ensuring $C^0$ continuity at the frozen boundary.
    \item \textbf{Safety (Hard Constraint):}
    \begin{equation}
    d\big(\mathcal{B}_k(\tau)\oplus \Etrack,\; \mathcal{B}_j(\tau)\oplus \Etrack\big) \geq \delta_{safe}, \quad \forall k \neq j
    \end{equation}
    \item \textbf{Terminal Constraint:} $x_k(t_k + T) \in \mathcal{X}_f$ (Terminal invariant set).
\end{enumerate}

\section{Complexity and Communication Scaling}
\subsection{Local computational complexity}
Let $N_i:=|\mathcal{EA}_i|+|\mathcal{S}_i|$ denote the number of owned and shadowed agents participating in coordinator $\mathcal{N}_i$'s local optimization. A key benefit of the hierarchical decomposition is that the per-coordinator optimization scales with $N_i$ rather than the global $N$. For typical dense quadratic-program or sequential convex programming (SCP) implementations, the dominant per-iteration solve scales as $O(N_i^3)$ in the worst case (dense linear algebra), while neighbor sparsity reduces this cost in practice.

Crucially, the spatial decomposition allows $N_i$ to be bounded by design. By splitting subspaces $\mathcal{W}_i$ until they are small enough (yet large enough to avoid handover thrashing, i.e., diameter $\gg v_{max}t_{step}$), and enforcing a fixed upper bound on the planning horizon $t_{planning}+t_{step}$, we guarantee that the computation task of each coordinator remains within the capacity of local hardware, regardless of the total system size $N$.

\subsection{Reliability and Redundancy}
To eliminate single-points-of-failure (SPoF) within a subspace, the architecture supports **online backup coordinators**. A secondary node can shadow the state of $\mathcal{N}_i$ and seamlessly take over control if the primary fails, ensuring high availability critical for infrastructure deployment.

\medskip
\noindent\textbf{Coordination overhead.}
Prollect adds a lightweight look-ahead conflict check in $W_3$ and a safety projection layer. These components scale approximately with the number of local neighbor interactions, i.e., $O(N_i\,d_i)$ where $d_i$ is the average neighbor count under the interaction radius.

\subsection{Communication load}
Communication is organized on the coordinator graph $G_{comm}$ and occurs primarily through spatiotemporal tubes. Each coordinator $\mathcal{N}_i$ transmits $\mathcal{ST}_k$ only to adjacent coordinators whose subspaces intersect the tube. Therefore, the per-cycle message count is bounded by
\[
O\!\left(\sum_i \sum_{a_k\in\mathcal{EA}_i} |\{j:(i,j)\in\mathcal{E}_{comm},\ \mathcal{ST}_k\cap \mathcal{W}_j\neq\emptyset\}|\right),
\]
which is typically proportional to the boundary-crossing rate rather than $N$.

\medskip
\noindent\textbf{Empirical scaling.}
We report runtime per control call in the scalability study (Table \ref{tab:scaling}), which provides an implementation-level validation of the expected neighbor-driven growth.

\begin{table*}[t]
\caption{Per-cycle scaling summary (order-level)}
\centering
\scriptsize
\setlength{\tabcolsep}{4pt}
\renewcommand{\arraystretch}{1.05}
\begin{tabular}{@{}lp{4.9cm}p{2.6cm}@{}}
\toprule
Component & Typical per-cycle cost & Driver \\ \midrule
Local solve (HVP) & $O(N_i^3)$ (dense worst case; often sparser in practice) & local agents + shadow agents \\
Look-ahead conflict check ($W_3$) & $O(N_i d_i)$ & neighbor interactions \\
Safety projection (velocity filter) & $O(N_i d_i)$ & local neighbor constraints \\
Tube exchange & $O(|\mathcal{E}_{comm}|\,\bar{s})$ & boundary crossings (tube size $\bar{s}$) \\
\bottomrule
\end{tabular}
\label{tab:complexity}
\end{table*}

\section{Stability Analysis}

We now provide formal feasibility and stability guarantees. The analysis relies on standard MPC terminal ingredients (existence of a terminal set/controller and a terminal decrease condition) and on the tube/tracking-envelope bridge (Assumption \ref{assump:track}, Proposition \ref{prop:tube_safety}) used to interpret ``hard safety'' under sampled execution. The boundary-coupling terms induced by Shadow Agents are handled via an ISS argument, which we use as a modular consistency property between adjacent coordinators.

\subsection{Terminal Ingredients (MPC Standard Assumptions)}
The following terminal ingredients are standard in MPC stability theory \cite{Rawlings2017MPC} and are used to obtain a value-function decrease.
\begin{assumption}[Terminal set and terminal controller]\label{assump:terminal}
There exist a compact terminal set $\mathcal{X}_f$ and a locally Lipschitz terminal feedback $\kappa_f:\mathcal{X}_f\to\mathcal{U}$ such that:
\begin{enumerate}
    \item (\textbf{Positive invariance}) If $x\in\mathcal{X}_f$, then the closed-loop trajectory under $u=\kappa_f(x)$ remains in $\mathcal{X}_f$ and satisfies all constraints, including the hard safety constraint (possibly in tightened/tube form).
    \item (\textbf{Terminal Lyapunov decrease}) There exists a continuous terminal cost $V_f:\mathcal{X}_f\to\R_{\ge 0}$ and a class-$\mathcal{K}_\infty$ function $\alpha_\ell$ such that, for the sampled dynamics over one update step,
    \begin{equation}
        V_f(x^+) - V_f(x) \le -\alpha_\ell(\|x-\sigma^{ref}\|), \qquad \forall x\in\mathcal{X}_f,
    \end{equation}
    where $x^+$ denotes the state after one update step under $u=\kappa_f(x)$.
\end{enumerate}
\end{assumption}

\begin{remark}[Concrete terminal ingredients for the planning abstraction]
For the velocity-command planning abstraction used to bridge to implementation (Section~\ref{subsec:model_mismatch}), one may take $\sigma^{ref}$ as a fixed goal configuration and choose $\mathcal{X}_f$ as a sufficiently small neighborhood of $\sigma^{ref}$ in which (i) input constraints are inactive (or handled by standard saturation arguments) and (ii) tightened safety constraints remain inactive due to a strict clearance margin. On such a neighborhood, standard quadratic terminal ingredients for (locally) Lipschitz dynamics yield a locally stabilizing $\kappa_f$ and $V_f$ satisfying the sampled Lyapunov decrease \cite{Rawlings2017MPC}. This remark explains why Assumption \ref{assump:terminal} is not restrictive for the coordination abstraction, while richer embodied models can be handled by reducing to a locally stabilizable tracking error system with a sufficiently small terminal neighborhood.
\end{remark}

\begin{assumption}[Stage cost bounds]\label{assump:stage}
For each agent stage cost $\ell_k(x_k,u_k)=\|x_k-\sigma_k^{ref}\|_Q^2+\|u_k\|_R^2$, there exists a class-$\mathcal{K}_\infty$ function $\alpha_\ell$ such that
\begin{equation}
\ell_k(x_k,u_k)\ge \alpha_\ell(\|x_k-\sigma_k^{ref}\|),
\end{equation}
and $\ell_k(x_k,u_k)=0$ iff $x_k=\sigma_k^{ref}$ and $u_k=0$.
\end{assumption}

\subsection{Recursive Feasibility}
Recursive feasibility ensures that if the system is safe at time $t_k$, there exists at least one valid control sequence at $t_{k+1}$ that maintains safety.

\begin{theorem}[Recursive Feasibility]\label{thm:rf}
Consider the Prollect optimization problem \eqref{eq:cost} with tube-inflated hard safety constraints (using $\Etrack$ as in Proposition \ref{prop:tube_safety}) and a terminal constraint $x_k(t_k+T)\in\mathcal{X}_f$. If the problem is feasible at time $t_k$, then it is feasible at $t_{k+1} = t_k + t_{step}$, provided $t_{step} \leq t_{frozen}$ and the executed motion follows the committed prefix within the tracking envelope of Assumption \ref{assump:track}.
\end{theorem}

\begin{proof}
See Appendix \ref{app:recursive}.
\end{proof}

\subsection{Boundary Consistency}
\begin{lemma}[Shadow Consistency]
For any agent $a_k$ shared by $\mathcal{N}_i$ and $\mathcal{N}_j$, the tracking error $e_k(\tau) = \|x_k^{(i)}(\tau) - x_k^{(j)}(\tau)\|$ converges to a bounded set $\Omega_\epsilon$ as $t \to \infty$.
\end{lemma}
\begin{proof}
This is an immediate corollary of the ISS result (Theorem \ref{thm:iss}). In particular, the disagreement dynamics are ISS with respect to a mismatch input $d(t)$ that aggregates solver discrepancy and packet-loss induced prediction error. Therefore $\|e_k(t)\|$ is ultimately bounded by a class-$\mathcal{K}$ function of $\sup_{s\in[0,t]}\|d(s)\|$; if $d(t)\to 0$ (e.g., consistent tube exchange and vanishing prediction mismatch), then $\|e_k(t)\|\to 0$.
\end{proof}

\subsection{ISS View of Shadow-Agent Coupling}
We formalize the boundary handover as an input-to-state stability (ISS) property, where the ``input'' captures local optimization mismatch and packet-loss induced prediction error.

\begin{definition}[ISS]\label{def:iss}
A system $\dot{e} = F(e,d)$ is ISS if there exist $\beta\in\KL$ and $\gamma\in\K$ such that
\begin{equation}
\norm{e(t)} \le \beta(\norm{e(0)},t) + \gamma\!\left(\sup_{s\in[0,t]}\norm{d(s)}\right),\qquad \forall t\ge 0.
\end{equation}
\end{definition}

\begin{theorem}[ISS of Shadow Consistency]\label{thm:iss}
Assume the local dynamics are Lipschitz with constant $L_f$ (Assumption 3) and the shadow coupling gain satisfies $\lambda_b > L_f$. Then the boundary disagreement dynamics admit an ISS-Lyapunov function and are ISS with respect to the mismatch input $d(t)$.
\end{theorem}
\begin{proof}
See Appendix \ref{app:iss}.
\end{proof}

\subsection{Asymptotic Stability}
\begin{theorem}[Asymptotic Stability]\label{thm:asym}
Assume fixed references $\sigma_k^{ref}$ (no further intent changes), recursive feasibility (Theorem \ref{thm:rf}), and Assumptions \ref{assump:terminal}--\ref{assump:stage}. Then the sampled closed-loop under Prollect admits the aggregated optimal value function $\mathcal{V}_k$ as a Lyapunov function at update times and is asymptotically stable with respect to the reference set (i.e., $x_k(t)\to \sigma_k^{ref}$ as $k\to\infty$).
\end{theorem}
\begin{proof}
Define the aggregated optimal value function at update times as
\begin{equation}
\mathcal{V}_k := \sum_i J_i^*(t_k),
\end{equation}
where $J_i^*(t_k)$ is the optimal cost of \eqref{eq:cost} for coordinator $i$ at time $t_k$. Under Assumptions \ref{assump:terminal}--\ref{assump:stage}, the standard ``shift-and-append'' candidate (Appendix \ref{app:recursive}) implies a one-step decrease bound of the MPC value function:
\begin{equation}
\mathcal{V}_{k+1} - \mathcal{V}_k \le - \sum_{a_k} \int_{t_k}^{t_{k+1}} \alpha_\ell(\|x_k(\tau)-\sigma_k^{ref}(\tau)\|)\, d\tau,
\end{equation}
which shows $\mathcal{V}_k$ is nonincreasing and bounded below. Summability of the right-hand side implies $\|x_k(\tau)-\sigma_k^{ref}(\tau)\|\to 0$ as $k\to\infty$ (standard MPC Lyapunov argument). Details are provided in Appendix \ref{app:asym_mpc}. \qed
\end{proof}

\section{Robustness to Communication Failures}
To address practical network imperfections, we model packet dropouts and connect the design of the Frozen Window to probabilistic safety.

\subsection{Bernoulli Packet Dropouts}
\begin{assumption}[Packet Dropout Model]\label{assump:drop}
We model the dissemination of coordination information for a given cycle as an i.i.d.\ Bernoulli \emph{broadcast blackout}: with probability $p_{drop}\in[0,1)$, the coordinator's packet (consensus/trajectory update for that cycle) is not delivered before it is needed, and agents execute their previously committed frozen plan. Blackouts are independent across cycles.
\end{assumption}

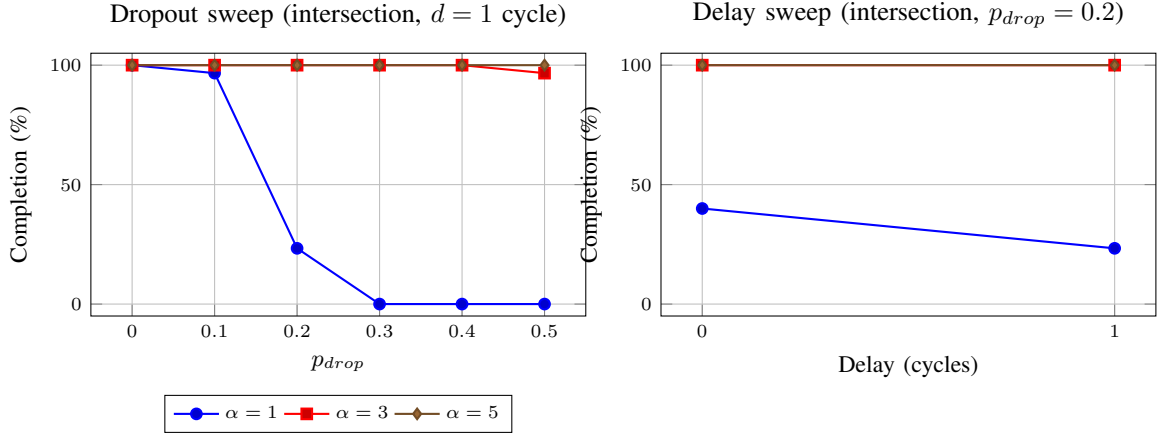
\begin{figure*}[t]
\centering
\begin{tikzpicture}
\begin{groupplot}[
    group style={group size=2 by 1, horizontal sep=1.0cm},
    width=0.45\textwidth,
    height=0.28\textwidth,
    grid=both,
    ymin=-5, ymax=105,
    legend style={font=\scriptsize, at={(0.5,-0.30)}, anchor=north, legend columns=3},
    tick label style={font=\scriptsize},
    label style={font=\small},
]
\nextgroupplot[
    title={Dropout sweep (intersection, $d=1$ cycle)},
    xlabel={$p_{drop}$}, ylabel={Completion (\%)},
]
\addplot+[mark=*, thick] table[x=p_drop,y=completion_rate_pct,col sep=comma] {sim/results/comm_plot_data/dropout_delay1_alpha1.csv};
\addlegendentry{$\alpha=1$}
\addplot+[mark=square*, thick] table[x=p_drop,y=completion_rate_pct,col sep=comma] {sim/results/comm_plot_data/dropout_delay1_alpha3.csv};
\addlegendentry{$\alpha=3$}
\addplot+[mark=diamond*, thick] table[x=p_drop,y=completion_rate_pct,col sep=comma] {sim/results/comm_plot_data/dropout_delay1_alpha5.csv};
\addlegendentry{$\alpha=5$}

\nextgroupplot[
    title={Delay sweep (intersection, $p_{drop}=0.2$)},
    xlabel={Delay (cycles)}, ylabel={Completion (\%)},
    xtick={0,1},
]
\addplot+[mark=*, thick] table[x=delay_steps,y=completion_rate_pct,col sep=comma] {sim/results/comm_plot_data/delay_p02_alpha1.csv};
\addplot+[mark=square*, thick] table[x=delay_steps,y=completion_rate_pct,col sep=comma] {sim/results/comm_plot_data/delay_p02_alpha3.csv};
\addplot+[mark=diamond*, thick] table[x=delay_steps,y=completion_rate_pct,col sep=comma] {sim/results/comm_plot_data/delay_p02_alpha5.csv};
\end{groupplot}
\end{tikzpicture}
\caption{Communication impairment sweep (intersection): completion versus blackout probability $p_{drop}$ (left) and delivery delay $d$ (right).}
\label{fig:comm_sweep}
\end{figure*}

\begin{remark}[Relation to link-wise dropouts]
Assumption \ref{assump:drop} matches the \emph{broadcast blackout} experiment in Fig.~\ref{fig:comm_sweep}. Link-wise i.i.d.\ dropouts on individual coordinator-to-coordinator or coordinator-to-agent links can be handled by analogous arguments by replacing the blackout probability with the probability of losing all required packets for a given agent over a horizon; we focus on the broadcast-blackout case because it directly captures the engineering requirement that the frozen buffer decouples execution from instantaneous message delivery.
\end{remark}

\begin{proposition}[Frozen-Window Safety Under Blackout]\label{prop:blackout}
If the Frozen Window spans $K_f=\lfloor t_{frozen}/t_{step}\rfloor$ cycles, then agents can execute a previously verified collision-free plan for at least $K_f$ consecutive cycles without receiving new coordination messages.
\end{proposition}

\begin{theorem}[Probabilistic Safety Design Rule]\label{thm:prob-safe}
Fix a target failure probability $\epsilon\in(0,1)$. Under Assumption \ref{assump:drop}, a sufficient design condition for ``blackout-safe'' execution with probability at least $1-\epsilon$ over a given horizon is:
\begin{equation}
K_f \ge \left\lceil \frac{\log(\epsilon)}{\log(p_{drop})}\right\rceil,\qquad p_{drop}\in(0,1),
\end{equation}
and if $p_{drop}=0$ then blackout-safe execution holds trivially.
\end{theorem}
\begin{proof}
Under Assumption \ref{assump:drop}, the probability of $K_f$ consecutive broadcast blackouts equals $p_{drop}^{K_f}$. Requiring $p_{drop}^{K_f}\le \epsilon$ yields the stated bound.
\end{proof}

\subsection{Interpretation and Practical Tuning}
The result shows why $t_{frozen}$ is not merely ``conservative'': it is an explicit safety buffer against communication jitter and outages. In practice, $p_{drop}$ can be estimated online and used to adapt $\alpha=t_{frozen}/t_{step}$ conservatively.

\subsection{Packet-dropout and delay sweep (numerical support)}
To numerically support the Bernoulli-dropout model (Assumption \ref{assump:drop}) and the blackout-safe interpretation (Proposition \ref{prop:blackout}), we conduct a communication-impairment sweep in the intersection benchmark. We model an i.i.d.\ Bernoulli \emph{broadcast blackout} per control cycle: with probability $p_{drop}$, the coordinator's packet is not delivered for that cycle, and agents continue executing their buffered frozen plan. We also model a fixed delivery delay of $d$ cycles, corresponding to $t_{tx}=d\,t_{step}$, and enforce the frozen-window compatibility by planning a horizon covering $t_{frozen}+t_{tx}$ (consistent with the design requirement $t_{frozen}\gg t_{tx}$).

\section{Numerical Validation}

\subsection{Simulation Setup}
We evaluate the \emph{efficiency-preserving} nature of Prollect: conflicts are detected \emph{before} they become physical, and the resulting velocity modifications are intentionally minor. To support large Monte Carlo sweeps efficiently, we implemented the simulator in C++17 and executed seeds in parallel. Agents are modeled as discs of radius $r=0.5m$ with unicycle-like execution of commanded planar velocity.

\medskip
\noindent\textbf{Prollect timing protocol (preemptive conflict check).}
At each coordinator cycle starting at time $t$, conflicts are detected in the \emph{detection interval}
\[
(t+t_{frozen}+t_{planning},\; t+t_{frozen}+t_{planning}+t_{step}],
\]
and, if found, the coordinator applies minor velocity adjustments over the \emph{adjustment interval}
\[
(t+t_{frozen},\; t+t_{frozen}+t_{planning}+t_{step}],
\]
so the system resolves conflicts \emph{before} they enter execution-critical time.

\begin{itemize}
    \item \textbf{Parameters:} $t_{step}=0.2s$, $t_{frozen}=0.2s$ ($\alpha=1$ unless otherwise stated), $t_{planning}=0.2s$, $t_{lookahead}=1.5s$, $v_{max}=1.5m/s$, integration $dt=0.05s$, max time $90s$, goal tolerance $2.0m$.
    \item \textbf{Baselines:}
    \begin{enumerate}
        \item \textbf{Reactive VO-projection:} a purely reactive projection-based correction inspired by the VO viewpoint \cite{Fiorini1998VO}: the intended command is straight-to-goal at $v_{max}$ and is then projected to satisfy short-horizon separation.
        \item \textbf{Reactive ORCA:} an ORCA-style half-plane formulation \cite{VanDenBerg2011ORCA} solved via 2D linear programming (reactive, reciprocal) with time horizon $T_{\text{ORCA}}=1.0$s.
        \item \textbf{DMPC-BR (iterative best-response MPC):} a lightweight DMPC-style controller that runs a small number of best-response iterations per cycle, using short-horizon neighbor trajectory prediction (constant-velocity over a 1.0s horizon) and sampling a finite set of constant-velocity candidates to minimize a goal-tracking plus separation penalty. This baseline has no frozen commitments and no proactive look-ahead window; it serves as a stronger distributed replanning comparator.
    \end{enumerate}
    \item \textbf{Implementation details (fairness):} all methods use the same neighbor selection radius ($20$m) and the same integration step ($dt=0.05$s). VO-projection denotes a purely reactive projection-based correction; ORCA uses the standard half-plane LP structure; DMPC-BR denotes iterative best-response replanning without frozen commitments or a proactive look-ahead window. All methods then pass their intended command through the same discrete-time safety projection operator $\Pi_{\text{safe}}$ (same implementation and numerical thresholding), and Prollect additionally logs when preemption is triggered.
    \item \textbf{Hard safety enforcement in discrete time (shared projection layer):} to match the paper's hard-safety requirement under sampled execution, commanded velocities are passed through a common discrete-time safety projection layer before integration (shared across baselines and Prollect). This layer enforces short-horizon separation at the command level. Therefore, to attribute performance \emph{beyond} this layer, we also report a \emph{projection activation rate} (ProjAct). Let $v^{int}(t_k)$ denote the intended command of a method before projection and $v^{exec}(t_k)$ the projected command that is executed; then
    \begin{equation}
        \mathrm{ProjAct} := \frac{1}{K}\sum_{k=1}^{K} \mathbf{1}\left\{\|v^{exec}(t_k)-v^{int}(t_k)\|>\eta\right\},
    \end{equation}
    where $\eta>0$ is a small numerical threshold (we use $\eta=10^{-9}$ in the simulator). Unless otherwise stated, the projection uses a fixed safety margin of $0.3$m and up to 6 projection iterations per call. A low ProjAct indicates that the method rarely requires last-moment safety correction.
    \item \textbf{Stand-off / non-termination (definition):} a run is counted as \emph{completed} iff all agents reach their goals within the maximum horizon $T_{\max}=90$s (the \textbf{Comp.} column in Tables~\ref{tab:intersection_eff}--\ref{tab:random_eff}). A run is said to exhibit a \emph{stand-off} (non-termination) if it does not complete by $T_{\max}$; we also log a \emph{deadlock} flag when the system fails to complete and the average speed falls below $0.1$\,m/s after the first $10$\,s, capturing a persistent stand-still mode. Thus, in our reported tables, stand-off rate is equivalently $1-\text{Comp.}$, and the deadlock flag is a stricter diagnostic of stand-off due to near-zero motion.
    \item \textbf{Reported statistics:} unless otherwise stated, we report the median [IQR] across 30 seeds, where IQR is the interquartile range (25th--75th percentile). Rates (completion/collision) are reported as fractions of runs.
    \item \textbf{Efficiency metrics:} completion time, average speed $\overline{\|v\|}$, and average velocity modification magnitude $\overline{\|\Delta v\|}=\overline{\|v-v_{nom}\|}$, where $v_{nom}$ points to the goal at $v_{max}$. Prollect also logs a preemption rate (fraction of control updates where a preemptive adjustment was triggered).
\end{itemize}

\subsection{Scenario A: 4-Way Intersection (Symmetry Stand-off)}
$N=20$ agents cross a 4-way intersection simultaneously. This benchmark is fully symmetric and is designed to expose reciprocal yielding stand-offs in purely reactive coordination. Under the stand-off definition above ($T_{\max}=90$s), the purely reactive baselines (VO-projection and ORCA) time out, corresponding to \textbf{Comp.=0\%} (100\% stand-off rate) in our Monte Carlo runs (Table \ref{tab:intersection_eff}). In contrast, both Prollect and the distributed replanning baseline DMPC-BR achieve 100\% completion. Prollect completes slightly faster and with smaller velocity disruption, while triggering preemption only on a small fraction of updates, illustrating the intended benefit of proactive look-ahead ($W_3$): resolving conflicts \emph{before} physical conflict with minor, sparse adjustments. Prollect also exhibits a lower projection activation rate (ProjAct) than the reactive VO-projection baseline, indicating reduced reliance on last-moment safety corrections (Table \ref{tab:intersection_eff}).

\subsection{Scenario B: Bidirectional Bottleneck (Throughput Under Constraints)}
We consider a narrow passage with bidirectional traffic ($N=16$). Prollect matches the reactive VO-projection baseline in completion reliability (100\%) and exhibits slightly smaller velocity modifications, confirming that conflicts are resolved early with minimal disruption. In this dense counterflow benchmark, reactive ORCA frequently exhibits reciprocal slow-down/oscillation and times out within the time limit (Table \ref{tab:bottleneck_eff}).

\subsection{Scenario C: Random Waypoint Navigation (Efficiency Preservation)}
$N=20$ agents navigate to random goals (non-overlapping starts). Prollect matches the reactive VO-projection baseline in completion reliability (96.7\%) and remains collision-free, while requiring slightly smaller velocity modifications on average. Reactive ORCA is also collision-free but has a lower completion rate and larger velocity modifications in this setting (Table \ref{tab:random_eff}).


\begin{table*}[t]
\caption{Intersection results (30 runs): median [IQR] where applicable}
\centering
\scriptsize
\setlength{\tabcolsep}{3pt}
\renewcommand{\arraystretch}{1.05}
\begin{tabular}{@{}lccccccc@{}}
\toprule
Method & Comp. & Coll. & Time (s) & Min Dist. (m) & $\overline{\Delta v}$ & Preempt & ProjAct \\ \midrule
Reactive VO-proj. & 0.0\% & 0.0\% & -- & 1.19 [1.19, 1.19] & 0.975 [0.975, 0.975] & 0.000 [0.000, 0.000] & 0.355 [0.355, 0.355] \\ 
Reactive ORCA & 0.0\% & 0.0\% & -- & 1.30 [1.30, 1.30] & 0.211 [0.211, 0.211] & 0.000 [0.000, 0.000] & 0.000 [0.000, 0.000] \\ 
DMPC-BR & 100.0\% & 0.0\% & 70.90 [70.90, 70.90] & 1.25 [1.25, 1.25] & 0.121 [0.121, 0.121] & 0.000 [0.000, 0.000] & 0.015 [0.015, 0.015] \\ 
Prollect & 100.0\% & 0.0\% & 69.25 [69.25, 69.25] & 1.22 [1.22, 1.22] & 0.066 [0.066, 0.066] & 0.063 [0.063, 0.063] & 0.033 [0.033, 0.033] \\ 
\bottomrule
\end{tabular}
\label{tab:intersection_eff}
\end{table*}

\begin{table*}[t]
\caption{Bottleneck results (30 runs): median [IQR]}
\centering
\scriptsize
\setlength{\tabcolsep}{3pt}
\renewcommand{\arraystretch}{1.05}
\begin{tabular}{@{}lccccccc@{}}
\toprule
Method & Comp. & Coll. & Time (s) & Min Dist. (m) & $\overline{\Delta v}$ & Preempt & ProjAct \\ \midrule
Reactive VO-proj. & 100.0\% & 0.0\% & 53.97 [53.67, 54.41] & 1.24 [1.22, 1.25] & 0.016 [0.010, 0.022] & 0.000 [0.000, 0.000] & 0.031 [0.020, 0.045] \\ 
Reactive ORCA & 3.3\% & 0.0\% & 72.65 [72.65, 72.65] & 1.28 [1.27, 1.30] & 0.510 [0.450, 0.548] & 0.000 [0.000, 0.000] & 0.007 [0.001, 0.015] \\ 
DMPC-BR & 100.0\% & 0.0\% & 53.82 [53.36, 54.52] & 1.27 [1.26, 1.27] & 0.039 [0.019, 0.053] & 0.000 [0.000, 0.000] & 0.006 [0.004, 0.009] \\ 
Prollect & 100.0\% & 0.0\% & 54.35 [54.07, 54.60] & 1.24 [1.23, 1.25] & 0.019 [0.013, 0.026] & 0.079 [0.047, 0.115] & 0.019 [0.013, 0.028] \\ 
\bottomrule
\end{tabular}
\label{tab:bottleneck_eff}
\end{table*}

\begin{table*}[t]
\caption{Random waypoint results (30 runs): median [IQR]}
\centering
\scriptsize
\setlength{\tabcolsep}{3pt}
\renewcommand{\arraystretch}{1.05}
\begin{tabular}{@{}lccccccc@{}}
\toprule
Method & Comp. & Coll. & Time (s) & Min Dist. (m) & $\overline{\Delta v}$ & Preempt & ProjAct \\ \midrule
Reactive VO-proj. & 96.7\% & 0.0\% & 52.75 [50.90, 56.95] & 1.24 [1.24, 1.25] & 0.017 [0.008, 0.031] & 0.000 [0.000, 0.000] & 0.014 [0.008, 0.020] \\ 
Reactive ORCA & 90.0\% & 0.0\% & 54.20 [51.32, 59.02] & 1.30 [1.29, 1.32] & 0.128 [0.087, 0.146] & 0.000 [0.000, 0.000] & 0.000 [0.000, 0.001] \\ 
DMPC-BR & 100.0\% & 0.0\% & 52.77 [50.99, 55.57] & 1.28 [1.27, 1.28] & 0.534 [0.467, 0.592] & 0.000 [0.000, 0.000] & 0.005 [0.002, 0.007] \\ 
Prollect & 96.7\% & 0.0\% & 52.75 [50.50, 56.95] & 1.24 [1.24, 1.26] & 0.020 [0.010, 0.030] & 0.039 [0.027, 0.049] & 0.014 [0.008, 0.019] \\ 
\bottomrule
\end{tabular}
\label{tab:random_eff}
\end{table*}

\noindent\textbf{Interpreting the new columns (attribution beyond safety projection).}
The \textbf{Preempt} column is the fraction of control updates where Prollect triggered a proactive adjustment in $W_3$ (preemption). The \textbf{ProjAct} column is the projection activation rate defined above: the fraction of control calls where the shared safety projection $\Pi_{\text{safe}}$ modified the intended command (with threshold $\eta$). Together, $\overline{\Delta v}$, Preempt, and ProjAct quantify whether a method resolves conflicts proactively (low Preempt/ProjAct with high completion) or relies on reactive last-moment corrections (high ProjAct).

\subsection{Parameter Sensitivity Analysis}
We report an ablation that isolates (i) the effect of preemptive look-ahead ($W_3$ enabled/disabled) and (ii) the frozen multiplier $\alpha=t_{frozen}/t_{step}$. Table \ref{tab:ablation} shows that disabling preemption destroys completion in the intersection benchmark (stand-off), while preemption yields 100\% completion with small $\overline{\Delta v}$. In these no-dropout experiments, varying $\alpha$ has little effect on completion or efficiency; its primary role is robustness under communication delay/dropout as formalized by Theorem \ref{thm:prob-safe}.

\begin{table*}[t]
\caption{Ablation study (parallel C++ simulator): effect of preemption ($W_3$ enabled) and frozen multiplier $\alpha=t_{frozen}/t_{step}$ on completion and efficiency (10 seeds). Each cell reports median runtime per control call (\,$\mu$s), median $\overline{\Delta v}$, median preemption rate, and completion rate (\%).}
\centering\scriptsize
\setlength{\tabcolsep}{3pt}\renewcommand{\arraystretch}{1.05}
\begin{tabular}{@{}lcccccc@{}}\toprule
Scenario & $\alpha$ & Preempt ON ($\mu$s/\,\,$\Delta v$/\,Pre/\,Comp) & Preempt OFF ($\mu$s/\,\,$\Delta v$/\,Pre/\,Comp) \\ \midrule
intersection & 1.0 & 1.95/0.065/0.064/100.0 & 2.58/0.975/0.000/0.0 \\ 
intersection & 2.0 & 1.95/0.065/0.064/100.0 & 2.58/0.975/0.000/0.0 \\ 
intersection & 3.0 & 1.95/0.065/0.064/100.0 & 2.58/0.975/0.000/0.0 \\ 
intersection & 5.0 & 1.95/0.065/0.064/100.0 & 2.58/0.975/0.000/0.0 \\ 
\midrule
random & 1.0 & 1.94/0.077/0.030/100.0 & 1.79/0.050/0.000/100.0 \\ 
random & 2.0 & 1.94/0.077/0.030/100.0 & 1.79/0.050/0.000/100.0 \\ 
random & 3.0 & 1.94/0.077/0.030/100.0 & 1.79/0.050/0.000/100.0 \\ 
random & 5.0 & 1.94/0.077/0.030/100.0 & 1.79/0.050/0.000/100.0 \\ 
\midrule
\bottomrule\end{tabular}
\label{tab:ablation}
\end{table*}

\subsection{Summary of comparative performance}
Across the evaluated scenarios, Prollect consistently preserves efficiency (small $\overline{\Delta v}$) while maintaining collision-free execution under the shared safety-projection layer. In the symmetry-sensitive intersection benchmark, purely reactive baselines (VO-projection and ORCA) exhibit 0\% completion (stand-off), while both Prollect and DMPC-BR complete; Prollect achieves comparable completion with smaller velocity disruption and sparse preemption. In the bottleneck benchmark, Prollect matches VO in completion and safety (both 100\% completion, 0 collisions) and keeps velocity modifications minor; reactive ORCA times out in this dense counterflow case. In the random waypoint benchmark, Prollect matches VO in completion reliability (both 96.7\%) with small velocity modification magnitude, while DMPC-BR completes reliably but can be substantially more disruptive in terms of $\overline{\Delta v}$ due to aggressive local replanning.

\subsection{Scalability Study}
To assess computational scalability, we vary the number of agents $N$ and report the median runtime per control call (in $\mu$s) together with completion rate, using 10 Monte Carlo seeds per configuration (generated and executed in parallel by the C++ simulator). Table \ref{tab:scaling} indicates the expected growth with $N$ due to neighbor interactions, and shows that Prollect maintains successful completion in the symmetry-sensitive intersection benchmark even at larger $N$, while incurring only a modest constant-factor overhead relative to the reactive VO baseline.

\begin{table*}[t]
\caption{Scaling study (parallel C++ simulator): median runtime per control call (\,$\mu$s) and completion rate (\%); each cell aggregates 10 Monte Carlo seeds.}
\centering\scriptsize
\setlength{\tabcolsep}{3pt}\renewcommand{\arraystretch}{1.05}
\begin{tabular}{@{}lcccccc@{}}\toprule
Scenario & $N$ & VO $\mu$s / Comp & ORCA $\mu$s / Comp & DMPC-BR $\mu$s / Comp & Prollect $\mu$s / Comp \\ \midrule
intersection & 20 & 2.43 / 0.0 & 1.99 / 0.0 & 109.28 / 100.0 & 2.32 / 100.0 \\ 
intersection & 40 & 3.80 / 0.0 & 3.53 / 0.0 & 183.71 / 100.0 & 4.66 / 100.0 \\ 
intersection & 80 & 6.84 / 0.0 & 5.16 / 0.0 & 218.08 / 100.0 & 5.10 / 100.0 \\ 
\midrule
bottleneck & 16 & 2.32 / 100.0 & 2.98 / 10.0 & 120.96 / 100.0 & 2.49 / 100.0 \\ 
bottleneck & 32 & 4.40 / 100.0 & 4.42 / 0.0 & 230.30 / 90.0 & 4.42 / 100.0 \\ 
bottleneck & 64 & 6.99 / 60.0 & 7.97 / 0.0 & 427.84 / 60.0 & 7.28 / 100.0 \\ 
\midrule
random & 20 & 1.59 / 100.0 & 1.84 / 90.0 & 72.15 / 100.0 & 1.87 / 100.0 \\ 
random & 40 & 2.94 / 100.0 & 3.38 / 60.0 & 144.19 / 100.0 & 3.24 / 100.0 \\ 
random & 80 & 5.46 / 50.0 & 6.52 / 0.0 & 288.80 / 100.0 & 6.04 / 60.0 \\ 
\bottomrule\end{tabular}
\label{tab:scaling}
\end{table*}

\begin{figure*}[t]
\centering
\begin{tikzpicture}
\begin{groupplot}[
    group style={group size=3 by 1, horizontal sep=0.9cm},
    width=0.31\textwidth,
    height=0.26\textwidth,
    grid=both,
    legend style={font=\scriptsize, at={(0.5,-0.28)}, anchor=north, legend columns=4},
    tick label style={font=\scriptsize},
    label style={font=\small},
    ymin=-5, ymax=105,
]
\nextgroupplot[title={Completion vs $N$ (intersection)}, xlabel={$N$}, ylabel={completion (\%)}]
\addplot+[mark=*, thick] table[x=n_agents,y=completion_rate_pct,col sep=comma] {sim/results/scaling_plot_data/intersection_vo.csv};
\addlegendentry{VO}
\addplot+[mark=square*, thick] table[x=n_agents,y=completion_rate_pct,col sep=comma] {sim/results/scaling_plot_data/intersection_orca.csv};
\addlegendentry{ORCA}
\addplot+[mark=triangle*, thick] table[x=n_agents,y=completion_rate_pct,col sep=comma] {sim/results/scaling_plot_data/intersection_dmpc.csv};
\addlegendentry{DMPC-BR}
\addplot+[mark=diamond*, thick] table[x=n_agents,y=completion_rate_pct,col sep=comma] {sim/results/scaling_plot_data/intersection_prollect.csv};
\addlegendentry{Prollect}

\nextgroupplot[title={Completion vs $N$ (bottleneck)}, xlabel={$N$}]
\addplot+[mark=*, thick] table[x=n_agents,y=completion_rate_pct,col sep=comma] {sim/results/scaling_plot_data/bottleneck_vo.csv};
\addplot+[mark=square*, thick] table[x=n_agents,y=completion_rate_pct,col sep=comma] {sim/results/scaling_plot_data/bottleneck_orca.csv};
\addplot+[mark=triangle*, thick] table[x=n_agents,y=completion_rate_pct,col sep=comma] {sim/results/scaling_plot_data/bottleneck_dmpc.csv};
\addplot+[mark=diamond*, thick] table[x=n_agents,y=completion_rate_pct,col sep=comma] {sim/results/scaling_plot_data/bottleneck_prollect.csv};

\nextgroupplot[title={Completion vs $N$ (random)}, xlabel={$N$}]
\addplot+[mark=*, thick] table[x=n_agents,y=completion_rate_pct,col sep=comma] {sim/results/scaling_plot_data/random_vo.csv};
\addplot+[mark=square*, thick] table[x=n_agents,y=completion_rate_pct,col sep=comma] {sim/results/scaling_plot_data/random_orca.csv};
\addplot+[mark=triangle*, thick] table[x=n_agents,y=completion_rate_pct,col sep=comma] {sim/results/scaling_plot_data/random_dmpc.csv};
\addplot+[mark=diamond*, thick] table[x=n_agents,y=completion_rate_pct,col sep=comma] {sim/results/scaling_plot_data/random_prollect.csv};
\end{groupplot}
\end{tikzpicture}
\caption{Completion rate scaling curves (10 seeds per point) extracted from \texttt{sim/results/scaling\_summary.csv}. These curves complement Table \ref{tab:scaling} by showing how reliability changes with density in symmetry-sensitive (intersection), throughput-limited (bottleneck), and unstructured (random) settings.}
\label{fig:scaling_completion}
\end{figure*}
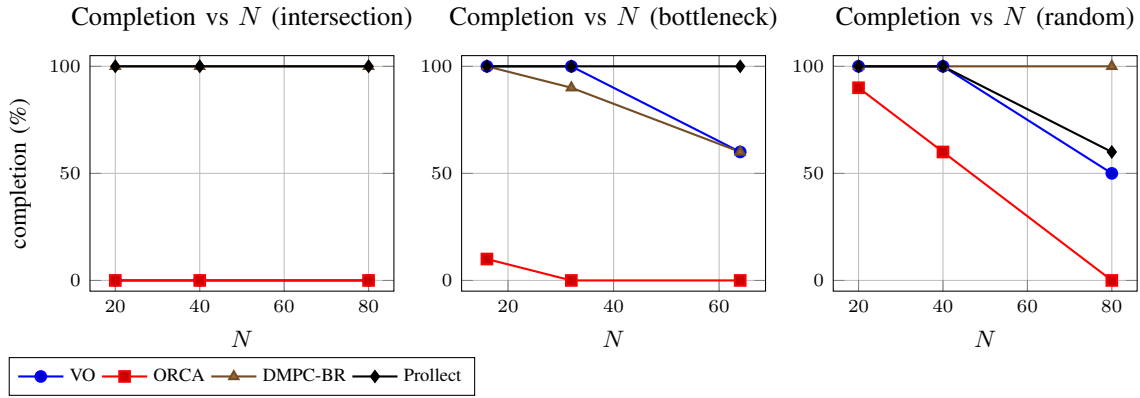

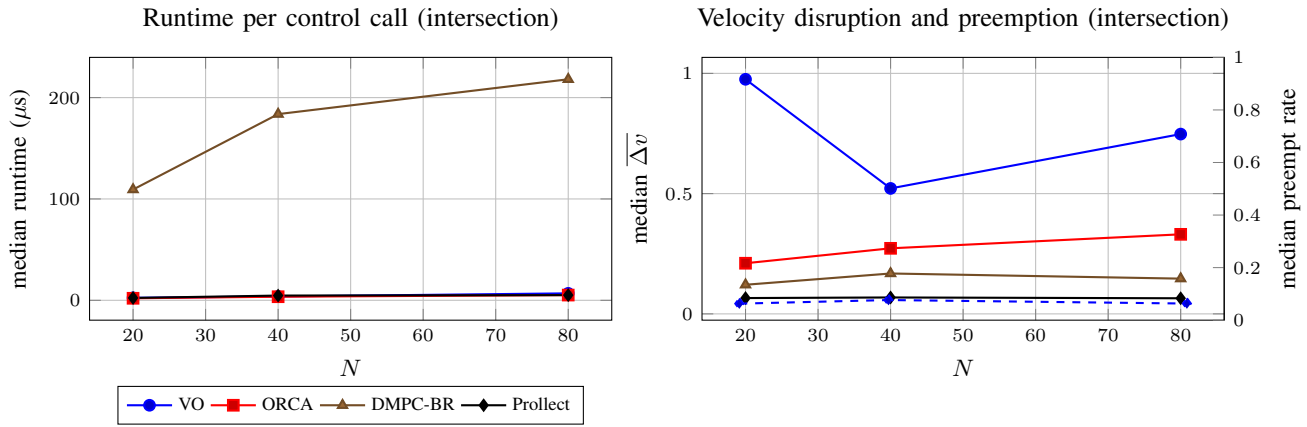
\begin{figure*}[t]
\centering
\begin{tikzpicture}
\begin{groupplot}[
    group style={group size=2 by 1, horizontal sep=1.2cm},
    width=0.47\textwidth,
    height=0.28\textwidth,
    grid=both,
    legend style={font=\scriptsize, at={(0.5,-0.25)}, anchor=north, legend columns=4},
    tick label style={font=\scriptsize},
    label style={font=\small},
]
\nextgroupplot[
    title={Runtime per control call (intersection)},
    xlabel={$N$}, ylabel={median runtime ($\mu$s)},
]
\addplot+[mark=*, thick] table[x=n_agents,y=runtime_us_med,col sep=comma] {sim/results/scaling_plot_data/intersection_vo.csv};
\addlegendentry{VO}
\addplot+[mark=square*, thick] table[x=n_agents,y=runtime_us_med,col sep=comma] {sim/results/scaling_plot_data/intersection_orca.csv};
\addlegendentry{ORCA}
\addplot+[mark=triangle*, thick] table[x=n_agents,y=runtime_us_med,col sep=comma] {sim/results/scaling_plot_data/intersection_dmpc.csv};
\addlegendentry{DMPC-BR}
\addplot+[mark=diamond*, thick] table[x=n_agents,y=runtime_us_med,col sep=comma] {sim/results/scaling_plot_data/intersection_prollect.csv};
\addlegendentry{Prollect}

\nextgroupplot[
    title={Velocity disruption and preemption (intersection)},
    name=disruptAxis,
    xlabel={$N$}, ylabel={median $\overline{\Delta v}$},
    yticklabel style={/pgf/number format/fixed},
]
\addplot+[mark=*, thick] table[x=n_agents,y=avg_dv_med,col sep=comma] {sim/results/scaling_plot_data/intersection_vo.csv};
\addplot+[mark=square*, thick] table[x=n_agents,y=avg_dv_med,col sep=comma] {sim/results/scaling_plot_data/intersection_orca.csv};
\addplot+[mark=triangle*, thick] table[x=n_agents,y=avg_dv_med,col sep=comma] {sim/results/scaling_plot_data/intersection_dmpc.csv};
\addplot+[mark=diamond*, thick] table[x=n_agents,y=avg_dv_med,col sep=comma] {sim/results/scaling_plot_data/intersection_prollect.csv};

\end{groupplot}

\begin{axis}[
    at={(disruptAxis.south west)},
    anchor=south west,
    width=0.47\textwidth,
    height=0.28\textwidth,
    xmin=15, xmax=85,
    axis x line=none,
    axis y line*=right,
    ylabel={median preempt rate},
    ymin=0, ymax=1,
    tick label style={font=\scriptsize},
    label style={font=\small},
]
\addplot+[mark=diamond*, dashed, thick] table[x=n_agents,y=preempt_rate_med,col sep=comma] {sim/results/scaling_plot_data/intersection_prollect.csv};
\end{axis}
\end{tikzpicture}
\caption{Scaling curves derived from \texttt{sim/results/scaling\_summary.csv} (10 seeds per point). Left: median runtime per control call versus number of agents. Right: median velocity disruption $\overline{\Delta v}$ (left y-axis) and Prollect preemption rate (right y-axis), illustrating that preemption remains sparse while maintaining low disruption.}
\label{fig:scaling_curves}
\end{figure*}

\noindent\textbf{Discussion of scaling curves.}
Figure \ref{fig:scaling_curves} complements Table \ref{tab:scaling} by revealing the \emph{trend} of implementation cost and disruption as $N$ increases. The runtime grows with neighbor interactions as expected, while remaining in the few-microsecond range per call in our C++ implementation. Importantly, Prollect's preemption rate stays low and does not explode with $N$ in the intersection benchmark, which supports the intended interpretation of Prollect as an \emph{efficiency-preserving} mechanism: the system resolves conflicts early but only triggers preemption on a small fraction of updates.

\medskip
\noindent On the disruption axis, the reactive VO-projection baseline shows large $\overline{\Delta v}$ in the symmetry-sensitive intersection setting (reflecting repeated reactive corrections without progress), whereas Prollect maintains a low and nearly $N$-independent $\overline{\Delta v}$ while still completing. This provides quantitative support for the paper's central claim: proactive look-ahead resolves conflicts \emph{before} physical conflict, so only minor velocity modifications are required.

\section{Discussion and Limitations}
\subsection{Why preemption improves throughput without sacrificing safety}
The intersection scenario highlights a key limitation of purely reactive coordination under full symmetry: reciprocal yielding can lead to non-termination (timeouts) even when hard safety is enforced. In our benchmark, reactive VO-projection and ORCA both time out under the $T_{\max}=90$s criterion (Table \ref{tab:intersection_eff}). Distributed replanning (DMPC-BR) can also break symmetry, but does so with larger velocity disruption than Prollect. Prollect breaks symmetry \emph{early} by using $W_3$ to detect conflicts before they become physical and to apply small, coordinated adjustments (low preemption rate and low $\overline{\Delta v}$), yielding reliable completion with minor disruption.

\subsection{Computation and communication overhead}
Relative to purely reactive baselines (VO/ORCA), Prollect incurs higher per-step computation because it performs look-ahead conflict checks and coordination logic. However, this overhead is bounded by the cycle constraint and is decoupled from safe execution by the Frozen Window. Communication is modeled as asynchronous/background transmission with delivery latency budget $t_{tx}$; the padding $t_{pad}>t_{tx}$ and snapshot/double-buffer rule prevent races between coordination dissemination and intent updates.

\subsection{Limitations and future work}
Our current evaluation uses simplified planar disc agents with unicycle-like execution and a small set of benchmark scenarios. The DMPC-BR baseline is a lightweight iterative best-response MPC and does not represent the full spectrum of advanced DMPC algorithms (e.g., ADMM/consensus MPC with robust tightening and formal convergence guarantees). Future work will include (i) richer vehicle models and obstacle maps, (ii) additional standardized distributed MPC baselines with explicit tuning protocols, (iii) communication models beyond i.i.d.\ Bernoulli drops (e.g., bursty losses and delays), and (iv) hardware/field experiments to validate the timing protocol and intent-update pipeline.

\subsection{Reproducibility}
All reported tables and plots are generated automatically by the parallel C++17 simulator (\texttt{sim\_cpp/run\_all.cpp}) and included via \texttt{\textbackslash input} from the \texttt{sim/results/} directory. For reviewer validation, the repository provides a one-shot script (\texttt{bash reproduce.sh}) that (i) rebuilds the simulator, (ii) regenerates all result artifacts in \texttt{sim/results/} (tables and plot-ready CSVs), and (iii) recompiles \texttt{manuscript.pdf}. Monte Carlo seeds are deterministic integers \texttt{0..mc-1}, and CSV outputs are written in a deterministic (sorted) order. Note that wall-clock runtime metrics (e.g., $\mu$s per call) depend on hardware and OS load; safety/completion/disruption metrics are deterministic given the seeds and simulator settings.

\section{Conclusion}
This paper presented the \textbf{Hierarchical Prollect Framework}, bridging the gap between theoretical stability and practical engineering. By introducing a robust timing protocol with mandatory idle buffers and a Shadow Agent handover mechanism, we demonstrated that large-scale MEA systems can be coordinated safely. The simulation results confirm that the "idle period" is not an inefficiency, but a necessary condition for stability in real-world distributed control.

\appendices

\section{Proof of Recursive Feasibility (Theorem \ref{thm:rf})}
\label{app:recursive}
Let $u^*(\cdot | t_k)$ be the optimal control trajectory computed at $t_k$, defined over $[t_k, t_k+T]$. We explicitly construct a candidate trajectory $\tilde{u}(\cdot | t_{k+1})$:

\textbf{1) Overlap Phase ($t_{k+1}$ to $t_k+T$):} Let $\tilde{u}(\tau) = u^*(\tau | t_k)$. This segment is a subset of the previously optimized trajectory. Since $u^*$ satisfied all hard safety constraints and dynamics in $W_2$ and $W_3$ at step $k$, it remains feasible when shifted into $W_1$ and $W_2$ at step $k+1$. Crucially, the segment entering the new Frozen Window $W_1$ was already safety-checked in the previous Planning Window $W_2$.

\textbf{2) Extension Phase ($t_k+T$ to $t_{k+1}+T$):} We apply a terminal control law $u_{term}(\tau) = \kappa_f(x(\tau))$ that renders the terminal set $\mathcal{X}_f$ invariant. Since $x^*(t_k+T) \in \mathcal{X}_f$ by the terminal constraint, the system remains inside $\mathcal{X}_f$.

The candidate $\tilde{u}$ satisfies dynamics, continuity, and safety. The Idle Buffer ensures that the computation of this candidate (or a better one) completes before the deadline. Thus, the feasible set is non-empty. \qed

\medskip
\noindent\textbf{Remark (Robust/Tube Variant).}
To connect the idealized continuous-time constraints to sampled implementation and bounded tracking error (Assumption \ref{assump:track}), one can adopt standard tube-based robust MPC ingredients \cite{Mayne2005TubeMPC}. Let the executed dynamics satisfy
\[
\dot{x}_k = f_k(x_k,u_k) + w_k,\qquad w_k(\tau)\in \Wdist,
\]
and let the nominal (planned) trajectory be $\bar{x}_k(\tau)$ with a tracking envelope $\Etrack$ so that $x_k(\tau)\in \bar{x}_k(\tau)\oplus \Etrack$. Then:
\begin{itemize}
    \item \textbf{Tightened constraints:} replace state constraints $\mathcal{X}_k$ by $\mathcal{X}_k\ominus \Etrack$, and enforce collision separation on inflated footprints as in \eqref{eq:cost} (tightened safety), i.e.,
    \[
    d\big(\mathcal{B}_k(\tau)\oplus \Etrack,\; \mathcal{B}_j(\tau)\oplus \Etrack\big)\ge \delta_{safe}.
    \]
    \item \textbf{Robust shift argument:} if the nominal solution at $t_k$ satisfies tightened constraints for all $\tau\in W_2\cup W_3$, then the shifted tail used at $t_{k+1}$ remains feasible under the same tightened constraints because it is a subset of the previously verified nominal plan. Proposition \ref{prop:tube_safety} then implies physical safety of the executed footprints.
    \item \textbf{Terminal robustness:} choose $\mathcal{X}_f$ as a (robust) positively invariant set for the nominal dynamics under $\kappa_f$, and ensure invariance under disturbance via standard tube constructions. This ensures the appended terminal segment remains feasible despite bounded perturbations.
\end{itemize}
This yields a fully robust version of Theorem \ref{thm:rf} under bounded disturbance/tracking error and aligns the proof with the hard-safety interpretation used in implementation.

\section{Proof of Zeno-Freeness (Theorem \ref{thm:zeno})}
\label{app:zeno}
We follow the standard definition of Zeno behavior in hybrid systems \cite{Goebel2012Hybrid}: an execution is Zeno if it exhibits infinitely many discrete transitions in finite continuous time.

\medskip
\noindent\textbf{Step 1 (One transition per cycle).}
By construction, the only transition that resets the timer is $q_{idle}\to q_{calc}$ with $\tau:=0$. The guard requires $\tau\ge t_{step}$, so each such transition consumes at least $t_{step}$ units of continuous time. Therefore, the sequence of reset times $\{t_k\}$ satisfies $t_{k+1}-t_k\ge t_{step}$, implying that there cannot be infinitely many reset transitions in finite time.

\medskip
\noindent\textbf{Step 2 (Positive dwell time in $q_{idle}$).}
The time spent in $q_{calc}$ is bounded by $t_{adj}^{max}$. The coordinator then remains in $q_{idle}$ until the end of the cycle at $\tau=t_{step}$. Thus, the duration in $q_{idle}$ is:
\begin{align}
    \Delta \tau_{idle} &= t_{step} - t_{adj} \nonumber
\end{align}
Substituting the design constraint $t_{step} > 1.5 t_{adj}^{max}$:
\begin{align}
    \Delta \tau_{idle} &> 1.5 t_{adj}^{max} - t_{adj}^{max} = 0.5 t_{adj}^{max} > 0
\end{align}
Since $\Delta \tau_{idle}$ is strictly bounded away from zero, the system cannot switch infinitely fast, preventing Zeno behavior. Furthermore, this positive dwell time absorbs computational jitter $\delta_{jitter}\in[0,0.5t_{adj}^{max})$.

\medskip
\noindent\textbf{Remark (Asynchronous transmission).}
If transmission is executed asynchronously in the background (overlapping with $q_{calc}$ and/or $q_{idle}$), it does not reduce $\Delta \tau_{idle}$ and therefore does not affect the Zeno/dwell-time argument.
Instead, communication enters as a delivery latency budget $t_{tx}$, which is handled by sizing the Frozen Window such that $t_{frozen} \gg t_{tx}$ (including jitter/outage margins). This decouples safe execution from instantaneous delivery. \qed

\medskip
\noindent\textbf{Remark (Why $t_{frozen} \gg t_{tx}$).}
Because agents execute the already-committed Frozen Window while messages are in transit (and transmission can be pipelined), the practical requirement is that dissemination completes well before the current frozen plan expires, i.e., $t_{frozen}$ should dominate $t_{tx}$ (including jitter/outage margins). This decouples safe execution from instantaneous communication and supports robust operation under network variability. \qed

\section{Proof of ISS of Shadow Consistency (Theorem \ref{thm:iss})}
\label{app:iss}
Let $e_k(t)=x_k^{(i)}(t)-x_k^{(j)}(t)$ denote the disagreement between coordinator $i$ and $j$ over a shared (shadowed) agent. Model the mismatch input as $d_k(t)$, capturing local solver discrepancy and packet-loss induced prediction error. Under Assumption 3 (Lipschitz dynamics), one can bound the disagreement dynamics by
\begin{equation}
\dot{e}_k = -\lambda_b e_k + \Delta_f(e_k) + d_k(t),
\end{equation}
where $\norm{\Delta_f(e_k)}\le L_f \norm{e_k}$. Consider the Lyapunov function $V(e_k)=\frac{1}{2}\norm{e_k}^2$. Then
\begin{align}
\dot{V} &= e_k^\top\dot{e}_k \le -\lambda_b\norm{e_k}^2 + L_f\norm{e_k}^2 + \norm{e_k}\norm{d_k} \nonumber\\
&=-(\lambda_b-L_f)\norm{e_k}^2 + \norm{e_k}\norm{d_k}.
\end{align}
For $\lambda_b>L_f$, complete the square to obtain
\begin{equation}
\dot{V} \le -\frac{\lambda_b-L_f}{2}\norm{e_k}^2 + \frac{1}{2(\lambda_b-L_f)}\norm{d_k}^2,
\end{equation}
which is an ISS-Lyapunov inequality. Therefore, the disagreement system is ISS (Definition \ref{def:iss}), with a gain that decreases as $\lambda_b$ increases. \qed

\medskip
\noindent\textbf{Explicit ISS bound (for completeness).}
From $\dot{V}\le -cV + \frac{1}{2(\lambda_b-L_f)}\|d_k\|^2$ with $c:=\lambda_b-L_f>0$ and $V=\frac{1}{2}\|e_k\|^2$, standard comparison arguments yield
\[
\|e_k(t)\|\le e^{-ct/2}\|e_k(0)\| + \sqrt{\tfrac{1}{c}}\ \sup_{s\in[0,t]}\|d_k(s)\|,
\]
which is of the form in Definition \ref{def:iss} with $\beta(r,t)=e^{-ct/2}r$ and $\gamma(r)=\sqrt{\tfrac{1}{c}}\,r$.

\section{Proof of Asymptotic Stability (Theorem \ref{thm:asym})}
\label{app:asym_mpc}
We provide a standard MPC Lyapunov proof adapted to the Prollect timing protocol.
\medskip

\noindent\textbf{Step 1 (Value function).}
Let $\mathcal{V}_k:=\sum_i J_i^*(t_k)$ denote the aggregated optimal cost at update time $t_k$.
Because $Q\succeq 0$, $R\succ 0$, and $V_f\ge 0$, we have $\mathcal{V}_k\ge 0$.

\medskip
\noindent\textbf{Step 2 (Shift-and-append candidate).}
Let $u^*(\cdot|t_k)$ be an optimal input trajectory at time $t_k$ for each coordinator. Construct a feasible candidate at $t_{k+1}=t_k+t_{step}$ by shifting the tail of $u^*(\cdot|t_k)$ over the overlap interval and appending the terminal controller $\kappa_f$ over the terminal segment, as in Appendix \ref{app:recursive}. Feasibility of the candidate follows from recursive feasibility (Theorem \ref{thm:rf}) and invariance of $\mathcal{X}_f$ under $\kappa_f$ (Assumption \ref{assump:terminal}).

\medskip
\noindent\textbf{Step 3 (One-step decrease).}
By optimality at $t_{k+1}$, the optimal cost is no larger than the cost of the candidate:
\begin{equation}
\mathcal{V}_{k+1}\le \mathcal{J}(\tilde{u}(\cdot|t_{k+1}),t_{k+1}).
\end{equation}
\noindent To make the cancellation explicit, write the aggregate cost at $t_k$ as
\begin{equation}
\begin{aligned}
\mathcal{V}_k
&=\sum_{a_k}\int_{t_k}^{t_k+T}\ell_k\big(x_k(\tau),u_k(\tau)\big)\,d\tau\\
&\quad + \sum_{a_k} V_f\big(x_k(t_k+T)\big).
\end{aligned}
\end{equation}
The shift-and-append candidate at $t_{k+1}$ uses the tail of the previous optimal input over $[t_{k+1},t_k+T]$ and the terminal controller over $[t_k+T,t_{k+1}+T]$. For compactness, define $\tilde{u}_{k+1}(\cdot):=\tilde{u}(\cdot|t_{k+1})$ and the terminal-control input as $u_{f,k}(\tau):=\kappa_f(x_k(\tau))$. Its cost therefore satisfies
\begin{align}
\mathcal{J}(\tilde{u}_{k+1},t_{k+1})
&=\sum_{a_k}\Big(
\int_{t_{k+1}}^{t_k+T}\ell_k(\cdot)\,d\tau \nonumber\\
&\qquad\qquad\quad
 + \int_{t_k+T}^{t_{k+1}+T}\ell_k\big(x_k(\tau),u_{f,k}(\tau)\big)\,d\tau \nonumber\\
&\qquad\qquad\quad
 + V_f\big(x_k(t_{k+1}+T)\big)\Big).
\end{align}
Subtracting $\mathcal{V}_k$ cancels the shared tail integral $\int_{t_{k+1}}^{t_k+T}\ell_k(\cdot)\,d\tau$ and yields the standard decrease structure (see also \cite{Rawlings2017MPC}):
\begin{equation}
\mathcal{V}_{k+1}-\mathcal{V}_k
\le -\sum_{a_k}\int_{t_k}^{t_{k+1}} \ell_k\big(x_k(\tau),u_k(\tau)\big)\,d\tau
+\Delta_f,
\end{equation}
where $\Delta_f := \sum_{a_k}\Big(V_f(x_k(t_{k+1}+T)) - V_f(x_k(t_k+T))\Big)$ is handled by the terminal Lyapunov decrease condition in Assumption \ref{assump:terminal}. In particular, on $\mathcal{X}_f$ we have
\begin{equation}
\Delta_f \le -\sum_{a_k}\int_{t_k+T}^{t_{k+1}+T}\ell_k\big(x_k(\tau),\kappa_f(x_k(\tau))\big)\,d\tau \le 0,
\end{equation}
and therefore
\begin{equation}
\mathcal{V}_{k+1}-\mathcal{V}_k
\le -\sum_{a_k}\int_{t_k}^{t_{k+1}} \ell_k\big(x_k(\tau),u_k(\tau)\big)\,d\tau.
\end{equation}
Using Assumption \ref{assump:stage}, $\ell_k(x_k,u_k)\ge \alpha_\ell(\|x_k-\sigma_k^{ref}\|)$, giving the decrease bound stated in Theorem \ref{thm:asym}.

\medskip
\noindent\textbf{Step 4 (Convergence).}
Because $\mathcal{V}_k\ge 0$ and is nonincreasing, it converges and the series $\sum_k \int_{t_k}^{t_{k+1}} \alpha_\ell(\|x_k(\tau)-\sigma_k^{ref}(\tau)\|)\, d\tau$ is finite. Since $\alpha_\ell$ is class-$\mathcal{K}_\infty$, this implies $\|x_k(\tau)-\sigma_k^{ref}(\tau)\|\to 0$ as $k\to\infty$ (standard MPC stability result; see \cite{Rawlings2017MPC}). \qed

\bibliographystyle{IEEEtran}
\bibliography{references}

@article{Li2024Preemptive,
  title={Preemptive Holistic Collaborative System and Its Application in Road Transportation},
  author={Li, Y. and Li, T. and Xu, X. and Dong, X. and Cai, Y. and Peng, T.},
  journal={arXiv preprint arXiv:2411.01918},
  year={2024}
}

@book{Pfeifer2006Body,
  title={How the Body Shapes the Way We Think: A New View of Intelligence},
  author={Pfeifer, R. and Bongard, J.},
  publisher={MIT Press},
  year={2006}
}

@book{Rawlings2017MPC,
  title={Model Predictive Control: Theory, Computation, and Design},
  author={Rawlings, J. B. and Mayne, D. Q. and Diehl, M.},
  publisher={Nob Hill Publishing},
  year={2017},
  edition={2nd}
}

@article{OlfatiSaber2007Consensus,
  title={Consensus and cooperation in networked multi-agent systems},
  author={Olfati-Saber, R. and Fax, J. A. and Murray, R. M.},
  journal={Proceedings of the IEEE},
  volume={95},
  number={1},
  pages={215--233},
  year={2007}
}

@article{Fiorini1998VO,
  title={Motion planning in dynamic environments using velocity obstacles},
  author={Fiorini, P. and Shiller, Z.},
  journal={The International Journal of Robotics Research},
  volume={17},
  number={7},
  pages={760--772},
  year={1998}
}

@article{VanDenBerg2011ORCA,
  title={Reciprocal n-body collision avoidance},
  author={van den Berg, J. and Guy, S. J. and Lin, M. and Manocha, D.},
  journal={Robotics Research},
  pages={3--19},
  year={2011},
  publisher={Springer}
}

@article{Mayne2005TubeMPC,
  title={Robust model predictive control: A survey},
  author={Mayne, D. Q. and Seron, M. M. and Rakovi{\'c}, S. V.},
  journal={Automatica},
  volume={41},
  number={5},
  pages={729--746},
  year={2005}
}

@book{Goebel2012Hybrid,
  title={Hybrid Dynamical Systems: Modeling, Stability, and Robustness},
  author={Goebel, R. and Sanfelice, R. G. and Teel, A. R.},
  publisher={Princeton University Press},
  year={2012}
}

@article{Wabersich2021PCBF,
  title={Predictive Control Barrier Functions: Enhanced Safety Mechanisms for Learning-Based Control},
  author={Wabersich, K. P. and Zeilinger, M. N.},
  journal={IEEE Transactions on Automatic Control},
  volume={68},
  number={5},
  pages={2638--2651},
  year={2023},
  doi={10.1109/TAC.2022.9779571}
}

@article{Ames2019CBF,
  title={Control Barrier Functions: Theory and Applications},
  author={Ames, Aaron D. and Coogan, Samuel and Egerstedt, Magnus and Notomista, Gennaro and Sreenath, Koushil and Tabuada, Paulo},
  journal={arXiv preprint arXiv:1903.11199},
  year={2019},
  month={3}
}

@article{Riccardi2025Partitioning,
  title={Partitioning Techniques for Non-Centralized Predictive Control: A Systematic Review and Novel Theoretical Insights},
  author={Riccardi, Alessandro and Laurenti, Luca and De Schutter, Bart},
  journal={arXiv preprint arXiv:2509.11470},
  year={2025},
  month={9}
}

@article{Kohler2019InexactDualDMPC,
  title={Distributed model predictive control---Recursive feasibility under inexact dual optimization},
  author={K{\"o}hler, Johannes and M{\"u}ller, Matthias A. and Allg{\"o}wer, Frank},
  journal={Automatica},
  volume={102},
  pages={1--9},
  year={2019},
  doi={10.1016/j.automatica.2018.12.037}
}

@article{Li2021DMPCCommNoise,
  title={Distributed model predictive control for linear systems under communication noise: Algorithm, theory and implementation},
  author={Li, Huiping and Jin, Bo and Yan, Weisheng},
  journal={Automatica},
  volume={125},
  pages={109422},
  year={2021},
  doi={10.1016/j.automatica.2020.109422}
}

@article{AlvesDoSanto2024AvoidMPC,
  title={Set-point tracking {MPC} with avoidance features},
  author={Alves do Santo, Marcelo and Ferramosca, Antonio and Raffo, Guilherme Vianna},
  journal={Automatica},
  volume={159},
  pages={111390},
  year={2024},
  doi={10.1016/j.automatica.2023.111390}
}

@article{Verginis2021AdaptiveAvoid,
  title={Adaptive robot navigation with collision avoidance subject to 2nd-order uncertain dynamics},
  author={Verginis, Christos K. and Dimarogonas, Dimos V.},
  journal={Automatica},
  volume={123},
  pages={109303},
  year={2021},
  doi={10.1016/j.automatica.2020.109303}
}

@article{Cao2025SMPCAvoid,
  title={Collision Avoidance Based on Stochastic Model Predictive Control in Collaboration Between {ROV} and {AUV}},
  author={Cao, Xiang and Wang, Xuerao and Sun, Changyin},
  journal={IEEE Transactions on Intelligent Transportation Systems},
  volume={26},
  number={7},
  pages={9461--9474},
  year={2025},
  doi={10.1109/TITS.2025.3562204}
}

@article{Shorinwa2024SeparableDMPC,
  title={Distributed Model Predictive Control via Separable Optimization in Multiagent Networks},
  author={Shorinwa, Ola and Schwager, Mac},
  journal={IEEE Transactions on Automatic Control},
  volume={69},
  number={1},
  pages={230--245},
  year={2024},
  doi={10.1109/TAC.2023.3269338}
}

@article{Mallick2025SwitchADMM,
  title={Distributed Model Predictive Control for Piecewise Affine Systems Based on Switching {ADMM}},
  author={Mallick, Samuel and Dabiri, Azita and De Schutter, Bart},
  journal={IEEE Transactions on Automatic Control},
  volume={70},
  number={6},
  pages={3727--3741},
  year={2025},
  doi={10.1109/TAC.2024.3512334}
}

@article{Ahn2018IntersectionSafety,
  title={Safety Verification and Control for Collision Avoidance at Road Intersections},
  author={Ahn, Heejin and Del Vecchio, Domitilla},
  journal={IEEE Transactions on Automatic Control},
  volume={63},
  number={3},
  pages={630--642},
  year={2018},
  doi={10.1109/TAC.2017.2729661}
}

@article{Li2022TubeContainment,
  title={Tube-based model predictive full containment control for stochastic multi-agent systems},
  author={Li, Liya and Shi, Peng and Ahn, Choon Ki and Kim, Yeong Jun and Xing, Wen},
  journal={IEEE Transactions on Automatic Control},
  pages={1--13},
  year={2022},
  doi={10.1109/TAC.2022.3202985}
}

\end{document}